\documentclass[prd, onecolumn, nofootinbib, floatfix,notitlepage,superscriptaddress]{revtex4-1}

\usepackage{amsmath}
\usepackage{graphicx}
\usepackage{dcolumn}
\usepackage{bm}
\usepackage{epsfig}
\usepackage{amssymb,latexsym,mathrsfs}
\usepackage{graphicx}
\usepackage{color,xcolor}
\usepackage{subfigure,float}
\usepackage{multirow}
\usepackage[normalem]{ulem}
\usepackage{hyperref}
\usepackage{ulem}
\usepackage{verbatim}
\hypersetup{
    colorlinks=true,
    linkcolor=red,
    citecolor=blue,
} 

\newcommand{\be}{\begin{equation}}
\newcommand{\ee}{\end{equation}}
\newcommand{\bs}{\begin{split}} 
\newcommand{\bea}{\begin{eqnarray}}
\newcommand{\eea}{\end{eqnarray}}

\def \mnras {Mon. Not. Roy. Astron. Soc.}
\def \aap {Astron. \& Astrophys.}
\def \apjl {Astrophys. J. Lett.}
\def \jcap {JCAP}

\newcommand{\tmin}{\dt_{\rm fit,min}}

\def \mubf {\mu_{\rm fit}}
\def \dtbf {\Delta t_{\rm fit}}
\def \mut {\mu_{\rm true}}
\def \dtt {\Delta t_{\rm true}}
\def \dt{\Delta t} 
\def \tsj{t_{s,j}}

\def \F{\mathcal{F}}

\def \czj {C_{0,j}}

\def \Fm {{F_j}^{\rm max}}

\begin{document}

\title{Be It Unresolved: Measuring Time Delays from Lensed Supernovae} 


\author{Satadru Bag}
\email{satadru@kasi.re.kr}
\affiliation{Korea Astronomy and Space Science Institute, Daejeon 34055, Korea}

\author{Alex G.~Kim}
\affiliation{Lawrence Berkeley National Laboratory, Berkeley, CA 94720, USA}

\author{Eric V.~Linder}
\affiliation{Lawrence Berkeley National Laboratory, Berkeley, CA 94720, USA}
\affiliation{Berkeley Center for Cosmological Physics, University of California, Berkeley, CA 94720, USA}
\affiliation{Energetic Cosmos Laboratory, Nazarbayev University, Nur-Sultan 010000, Kazakhstan}

\author{Arman Shafieloo}
\affiliation{Korea Astronomy and Space Science Institute, Daejeon 34055, Korea}
\affiliation{University of Science and Technology, Daejeon 34113, Korea}

\begin{abstract} 
Gravitationally lensed Type Ia supernovae may be the next frontier in cosmic 
probes, able to deliver independent constraints on dark energy, spatial curvature, and the Hubble constant. Measurements of time delays between the multiple images become more 
incisive due to the standardized candle nature of the source, monitoring for months rather than years, and partial 
immunity to microlensing. While currently 
extremely rare, hundreds of such 
systems should be detected by upcoming time domain surveys. Even more will have the images spatially unresolved, with the observed lightcurve a superposition of time delayed image fluxes. We investigate whether 
unresolved images can be recognized as lensed sources given only lightcurve 
information, and whether time delays can be extracted robustly. We develop 
a method that successfully identifies such systems, with a false positive 
rate of $\lesssim5\%$, and measures the time delays with a completeness of $\gtrsim93\%$ and with a bias of $\lesssim0.5\%$ for 
$\dtbf\gtrsim10$ days.  
\end{abstract}

\date{\today} 

\maketitle

\section{Introduction}

Gravitational lensing is the most visually striking demonstration 
of the curvature of spacetime. An individual source emits signals that 
reach the observer from different directions and at different times, 
forming multiple images. Moreover, these image separations and time 
delays carry important cosmological information on the distances and  
gravitational potentials along the lines of sight. For a variable source, repeated monitoring of the image flux allows estimation of the time delay, an important ingredient for the time delay distance  mapping the cosmic expansion. 

Time delay distances derived from monitoring quasars over years or even a decade are in use as a cosmological probe 
\cite{Wong:2019kwg,Millon:2019slk,Shajib:2019toy,Treu:2016ljm}. Quasars are plentiful, but their intrinsic 
time variability is unpredictable and it is challenging to separate out 
microlensing -- additional time variation due to motion of objects near 
the lines of sights. This means that often several years ($\gtrsim5$) 
of data must be accumulated before time delays can be measured with 
precision. With forthcoming surveys, detections of lensed Type 
Ia supernovae (SNe) -- supernovae were the 
source \citet{1964MNRAS.128..307R} originally proposed for the time delay cosmography technique -- will become much more numerous (only two systems 
have currently been published \cite{2015Sci...347.1123K, 2017Sci...356..291G}). These have the 
advantage of possessing better known time behavior (i.e.\ their flux vs 
time, or lightcurves, are more standard), with a characteristic time scale 
of a couple of months, enabling more rapid time delay estimation, and 
have a period of insensitivity to microlensing when measured in multiple 
wavelength filters \cite{2018ApJ...855...22G}. 

Supernova time delays have been put forward as ways to 
measure the Hubble constant \citep{2010MNRAS.405.2579O}, with application to 
the first lensed SN, called Refsdal, discussed in detail in  \cite{2015Sci...347.1123K, 2016ApJ...831..205K, 2016ApJ...820...50R, 2018ApJ...860...94G, 2018ApJ...853L..31V}. This involves cluster lensing 
with clearly spatially separated images. Simulation studies 
of lensed SN with images brightening well separated in time, 
i.e.\ time delays of $\gtrsim60$ days, is discussed in  
\cite{2019ApJ...876..107P}. Here  we concentrate on the opposite 
regime, where images are neither spatially nor temporally 
separated -- unresolved. 

Since typical galaxy lensing time delays are less than or of order the SN 
rise and fall time, when the images are not spatially well separated 
the observation is of an overlapping, summed lightcurve, offering a challenge 
to the time delay measurement. Indeed, one expects more lensed SNe to be 
unresolved than resolved. 
Here we take up the challenge and explore whether we can 1) recognize a  
SN lightcurve as being composed of multiple unresolved images, and 
2) accurately measure the time delay. 
Due to their property of having somewhat standard lightcurve shapes, and that 
upcoming surveys will devote considerable resources to studying them, including through follow up observations, we focus on Type Ia supernovae; however, our fitting approach 
is suitable for a wide variety of transients. This focus also follows from the use of Type Ia SN as luminosity distance indicators in addition; the combination of luminosity distance and time delay distance (possibly from two different SN Ia samples) can be quite powerful for a variety of cosmological characteristics, e.g.\ \cite{2018JCAP...03..041D,2020ApJ...895L..29L,Lyu:2020lwm,Pandey:2019yic,Ma:2019mwy,Arendse:2019hev,Liao:2019qoc,Taubenberger:2019qna,Collett:2019hrr,Liao:2017yeq,Treu:2016ljm}. 

In Section \ref{sec:data} we discuss the data sets expected from upcoming 
surveys, including the multiband lightcurve and noise properties. We introduce 
our lightcurve fitting method in Sec.~\ref{sec:method}, and test it in several ways on
mock data characteristic of upcoming surveys in Sec.~\ref{sec:test}. A summary and avenues for future 
development are presented in Sec.~\ref{sec:concl}.

\section{Forthcoming Supernova Datasets} \label{sec:data} 
High-redshift supernovae are efficiently discovered with
cadenced ``rolling'' wide-field imaging surveys \citep{2006A&A...447...31A}.  Fields of view of
tens of arcminutes and greater ensure that multiple SNe are active
within an untargeted pointing.  A cadence of several days ensures that
those supernovae will be discovered, ideally early in their evolution
to ensure follow-up observations.  The survey telescope, instrumentation,
and design define the search volume, both in terms of the monitored solid angle
and the depth, which then determine the number of supernovae that
can be discovered.

Gravitationally lensed SNe are relatively rare, only one in roughly a thousand  SNe are
expected to be strongly lensed.  Only surveys that monitor a large volume over a long
control time are capable of discovering interesting numbers of lensed SNe.
The Zwicky Transient Facility (ZTF) \citep{Bellm_2018} and the Vera C.\ Rubin Observatory Legacy Survey of 
Space and Time (LSST) \citep{2019ApJ...873..111I} are two surveys that will discover an interesting
number of gravitationally-lensed SNe of all types.

ZTF is a three year survey and is
composed of a public survey, which monitors 15,000~deg$^{2}$ every 3 nights to $g, r \approx 20.5$~mag, 
a partnership survey that adds supplemental high-cadence $g, r$ and $i$ exposures for a subset of sky,
and Caltech time.  The nominal LSST wide-fast-deep (WFD) 10-year survey (known as {\tt minion 1016})
covers $\sim$20,000~deg$^{2}$ and cycles $ugrizy$ every two-three weeks to a limiting magnitude
of $\sim 23.5$~mag.
\citet{Goldstein:2018bue} forecast that ZTF (LSST) can make early discoveries of 0.02 (0.79) 91bg-like, 0.17 (5.92) 91T-like, 1.22 (47.84) Type Ia, 2.76 (88.51) Type IIP, 0.31 (12.78) Type IIL, and 0.36 (15.43) Type Ib/c lensed SNe per year. They also forecast that the surveys can discover at least 3.75 (209.32) Type IIn lensed SNe per year, for a total of at least 8.60 (380.60) lensed supernovae per year under fiducial observing strategies.
The simulations and resultant light curves used to make these forecasts serve as
the
test 
data used in this article. 
We concentrate on the ZTF simulations in this work, to get an indication of nearer term results. The method  however applies equally  well to LSST simulation data. 
See also \cite{Wojtak:2019hsc,Huber:2019ljb,2019arXiv190205141V,2018ApJ...855...22G}. 

For lens systems with small Einstein radii,
image separations of multiple images can be too small
to be spatially resolved in observations.
\citet{Goldstein:2018bue} find that
over half of strongly-lensed SNe discovered by LSST will have a minimum angular
separation of $<1''$ and a significant fraction of those will have separations of $<0.5''$.
Many strongly-lensed supernova discoveries
will thus be unresolved by traditional ground-based
observations.
Robust determination of
time delays from the single observed light curve comprised of the sum of light from multiple unresolved images
could be valuable in  increasing the number of lensed supernova systems usable for cosmology from upcoming surveys.

Supplemental high-resolution imaging, e.g.\ from adaptive optics or space observations,
can deliver important information. For example, an external determination
of the number of images in the lensed system, and hence the number
of time delays, fixes what otherwise would be a free parameter
in the fit of an unresolved light curve. However, this would likely be at a single epoch, and the monitoring telescope that delivers the lightcurve might not separate the images. Thus the technique presented here is still of use in these cases.

\section{Fitting an Unresolved Lightcurve} \label{sec:method}

\subsection{Modeling the Lightcurve} 

The observed unresolved lightcurve is the sum of the fluxes from each of the unresolved images, each image $i$ having the same light curve shape since it comes from the same source, but with differing amplitudes
$a_i$
and delayed observed phases, i.e.\ dates of explosion $t_i$. 

The light curve observed in the $j$th filter (we use filter and band interchangeably), summed over $N_I$ images, can be written as 
\be \label{eq:F}
F_j(t) =\sum_{i=1}^{N_I} a_i\,f_j(t-t_i) \,,
\ee 
where $f_j$ is the intrinsic light curve in the $j$th filter.
As the science measurements of interest are the relative time 
delays and magnifications, we work 
in terms of the parameters 
$\Delta t_i \equiv t_i-t_1$ and $\mu_i \equiv a_i/a_1$   defined relative to the first, earliest image.  
For simplicity, one can choose $t_1=0$ 
so that Eq.~\eqref{eq:F} is 
\be\label{eq:F1}
F_j(t) =\sum_{i=1}^{N_I} \mu_i  \F_j(t-\Delta t_i)\,, 
\ee 
where of course $\mu_1=1$ and $\dt_1=0$, and $\F_j(t) \equiv a_1 f_j(t)$ is a scaled version of the source lightcurve. 

Our model fits both the underlying multi-band
light curve shapes, $\F_j(t)$, and the relative magnifications and
time delays of each of the images. Using a fixed template for $\F_j(t)$ is unlikely to 
give the flexibility for fits to a diverse set of supernovae, and indeed 
may lead to biases on time delay estimation, so we want 
to keep $\F_j(t)$ adaptable. The underlying multi-band light curve shapes $\F_j(t)$ are modeled as a fiducial function with a shape that is generically similar to supernovae, modified 
multiplicatively by a truncated series of orthogonal functions.
Specifically, we choose $\F_j(t)$ 
to be the product of a log-normal function multiplied by a basis expansion with respect to
the first four Chebyshev functions, 
\begin{align}\label{eq:template} 
\F_j(t) & = N_j \,\frac{1}{t} \exp\left[-\frac{(\ln t - b_j)^2}{2 \sigma_j^2}  \right] \\ 
&\qquad \times \left[1 + C_{1,j} \tsj + C_{2,j} (2\tsj ^2 -1) + C_{3,j} (4\tsj ^3 - 3 \tsj)+ C_{4,j} (8\tsj ^4 -8 \tsj ^2 +1) \right]\notag \;.
\end{align}  
Chebyshev functions have been widely used in Crossing statistics 
to give the desired flexibility 
\cite{Shafieloo:2010xm, Shafieloo:2012yh, Shafieloo:2012jb, Hazra:2014hma}. 
The time variable in the crossing terms uses 
$\tsj \equiv t/t_{j, {\rm max}} -1$\footnote{Supernovae are observed 
in modified Julian day (MJD). We subtract the time of the start of 
the observation in MJD from all the observation times. 
The quantity $t_{j,{\rm max}}$ is the time we end the fit in band $j$ (e.g.\ due to lack of data or fading of the supernova).
Note that a non zero $t_1$ in Eq.~\ref{eq:F} simply shifts the light curves earlier or later in time; we actually consider $t_1$ as a free parameter (same for all the filters) to allow the fit to the observed lightcurve data to begin a little before or after the nominal explosion time, to account for random scatter in the data near zero flux, so $t \to t - t_1$ in Eq.~\eqref{eq:F1}. 
}.  

In this model, the underlying
light curve in a single filter is described by one normalization $N_j$ and six shape parameters: $b_j$, $\sigma_j$, $C_{1,j}$,
$C_{2,j}$,
$C_{3,j}$,
$C_{4,j}$.
Thus to describe light curves composed of $N_I$ 
images in $N_B$ bands, there are $2(N_I-1)+7N_B+1$ parameters, where there is 1 parameter for the numerical fit shift parameter $t_1$. 

In this article we make two simplifying restrictions. We concentrate 
on systems with only two images (or simply 
one if it is unlensed, for testing). 
The unresolved light curve consisting of two images is given by 
\begin{equation}\label{eq:Ff}
    F_j(t)=\F_j(t)+\mu \, \F_j(t - \dt)\;,
\end{equation}
where the light curve of the first image $\F_j(t)$ is described by Eq.~\eqref{eq:template} (together with the shift parameter $t_1$) 
and the second image by a shifted, scaled version of it.
For the ZTF-like systems we have observations in $g,r,i$ filters.  Therefore, we have $2 + 7 \times 3 +1=24$ fit parameters, two of which are the parameters of cosmological interest, namely the relative time delay $\dt$ between the images and their relative magnification $\mu$. 
The second simplification is that here we do not include the effect of microlensing on the light curve (nor do the simulations we use 
from \citet{Goldstein:2018bue}). 
Microlensing can introduce generally small, slowly changing magnifications that
are different for each image, though correlated for the different filters of the same image. 
(Indeed, \citet{Goldstein:2018bue} have shown that 
the color, i.e.\ differences between filters, is insensitive to microlensing for much of the phase relevant to time delay 
estimation.)  
For the small image separations we consider here for unresolved 
lensed SN, microlensing effects between images may also be reduced. 
Nevertheless, expanding our model to include further multiplicity of 
images and microlensing is planned for future work.

\subsection{Sampling and setting the priors} 

We employ Hamiltonian Monte Carlo (HMC) for sampling, since it is well suited for dealing with a few dozen parameters. Specifically, we use the PyStan package \cite{pystan}, a python interface to STAN \cite{stan_2017}. 
We set priors on the parameters as described below, balancing broadness of the priors with convergence efficiency. 
As priors tend to involve the astrophysics of the source, their specific 
values need to be adjusted depending on the type of transient. 
Here we focus on normal Type Ia supernovae. Apart from the specific 
priors, however, the method described in the main text should be 
broadly applicable to many unresolved lensed transients.

\subsubsection{Prior on time delay and magnification} \label{sec:dtprior} 

\begin{figure}
\centering
\subfigure[\ System No=51127319, $\dt= 15.58$ days]{
\includegraphics[width=0.485\textwidth]{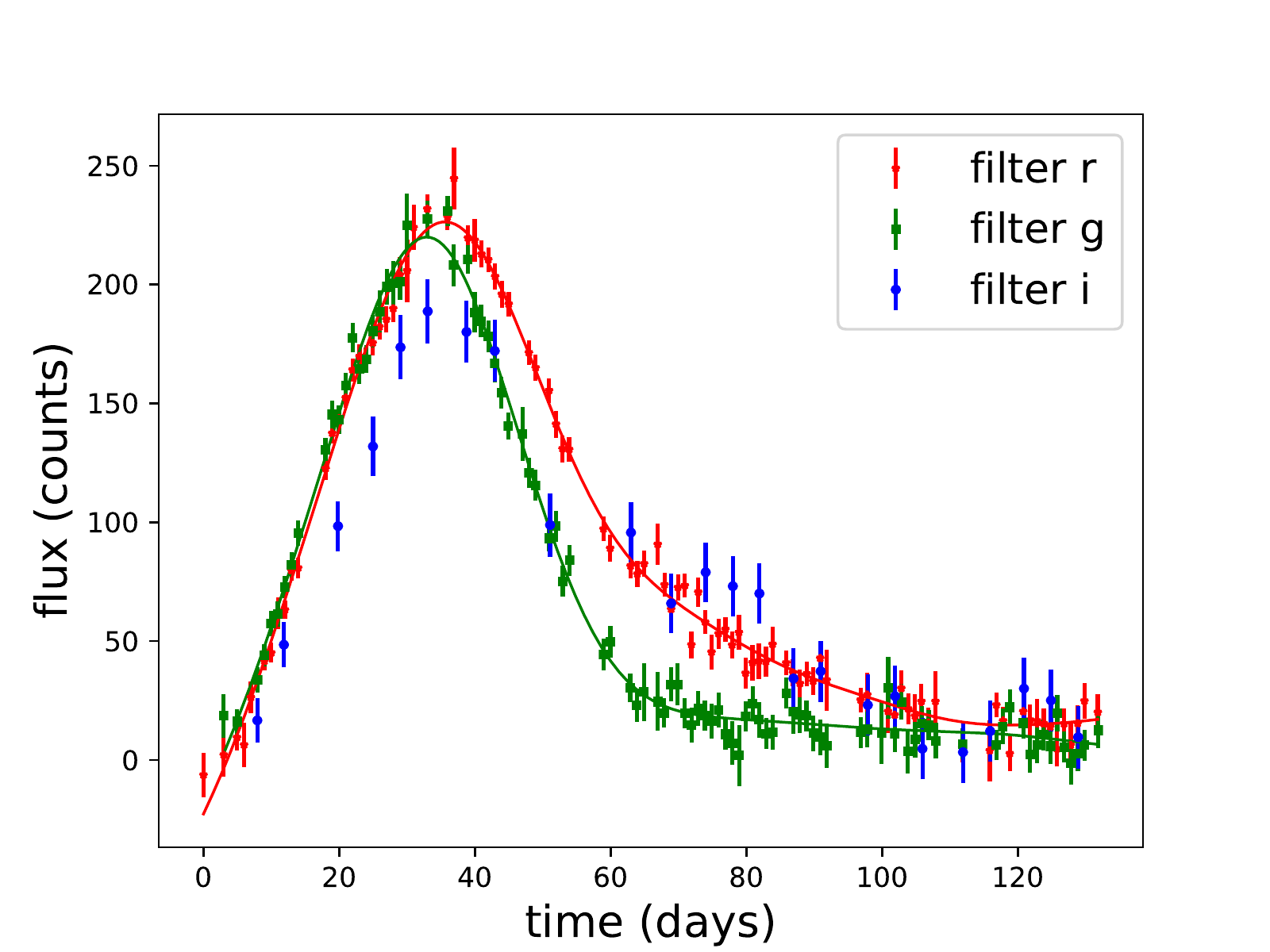}}
\subfigure[\ System No=40221141, $\dt= 42.10$ days]{
\includegraphics[width=0.485\textwidth]{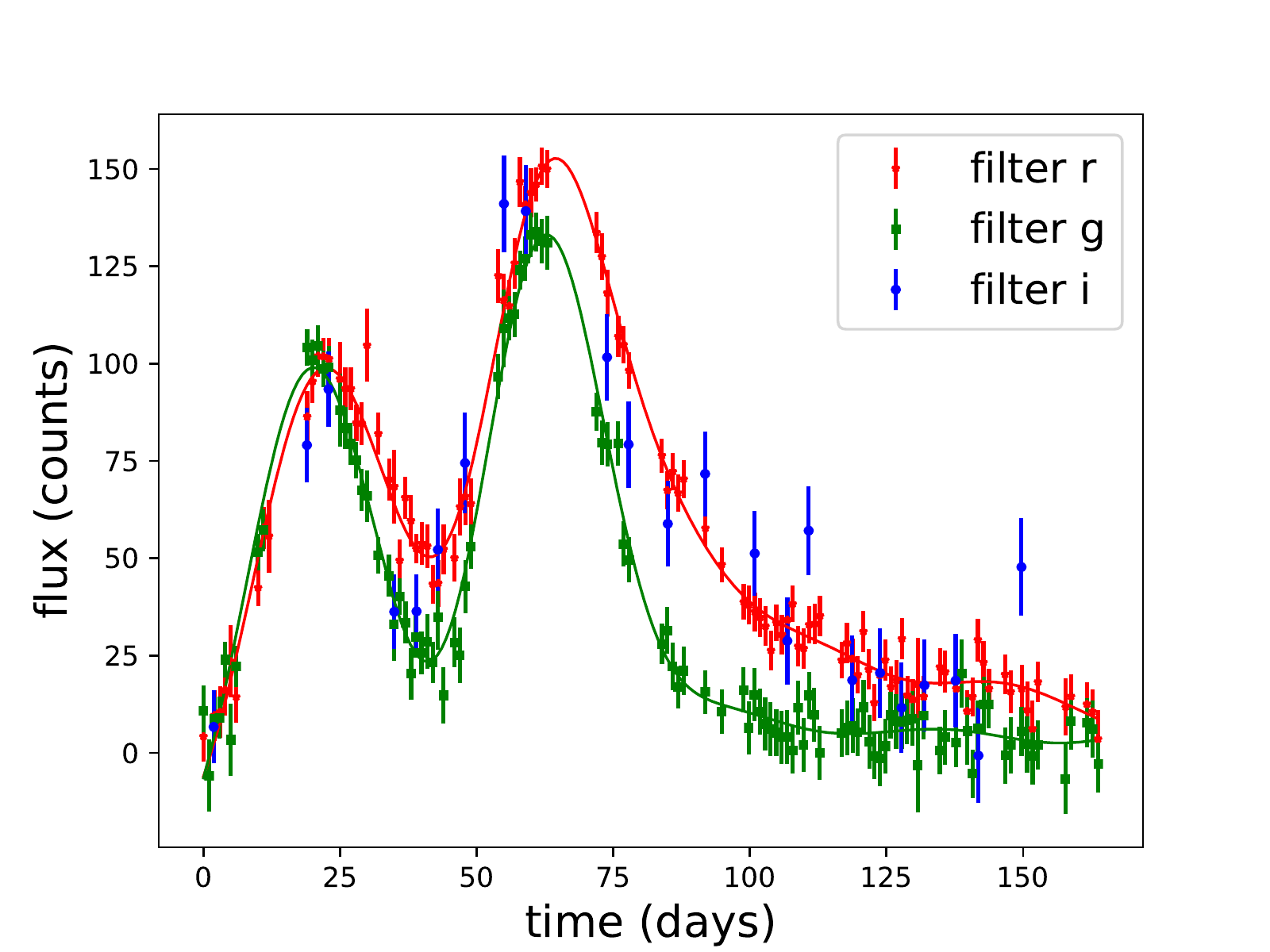}}
\caption{
Mock observations with ZTF characteristics are shown from the compilation of simulated Type-1a SNe \cite{Goldstein:2018bue} in 
$g$, $r$, $i$ filters. 
Solid curves show a smoothing of the data, 
helpful for identifying the number of peaks by eye.  The left panel, having a smaller time delay,  shows 
a single peak, while the right panel with a longer time delay shows two peaks corresponding to the two images. 
We do not use $i$ band data for determining the number of peaks due to its generally poorer data quality and longer decline time. 
}
\label{fig:smooth_lc}
\end{figure}

For time delays that are shorter than the characteristic source lightcurve width, the observed lightcurve of the combined images will tend to have one consolidated peak, while for longer time delays two separate peaks, one for each of the images, may be evident (depending on the magnification ratio). 
Long time delays and large magnification ratios would tend to 
give larger angular separation also, leading to 
resolved rather than unresolved lensed supernovae. However, we will apply our method 
to a wide range of time delays, though we concentrate on 
the more difficult, lower end of the range, $\Delta t\lesssim 30$. Examples of mock observed lightcurves of 
lensed supernovae are given in Fig.~\ref{fig:smooth_lc}.

As seen in Fig.~\ref{fig:smooth_lc}, 
when the peaks are well separated one can  roughly estimate the time delay from the peak positions and the relative magnification from the ratio of the peaks. 
These estimations can be used to set tighter priors for these large time delay systems. Conversely, when there is no evident double peak structure (in the filters we use), this imposes an upper limit on the time delay, based on the width of the peak. These 
features can be helpful in setting 
the priors on $\dt$ (and $\mu$ in some cases) so that the fit 
converges faster and we ameliorate 
issues of nonconvergence, or degenerate or catastrophically false fits. 

The procedure begins by identifying the peaks in the smoothed light curve data (this is solely for identifying the number of peaks; the full data is used in the fit). After testing different smoothing methods we use the iterative smoothing algorithm with exponential kernel in  \cite{Shafieloo:2005nd} (see also, \cite{Shafieloo:2007cs, Shafieloo:2009hi, Shafieloo:2012yh, Aghamousa:2014uya}), with a smoothing scale of $12$ days. While searching for peaks, to eliminate false peaks due to noise  we set a minimum height of the peaks relative to the tallest peak, minimum width, and minimum separation between peaks (details below). 

\begin{itemize} 
\item If $1$ peak identified: the time delay is not very high and we set a prior $0 \leq \dt \leq 30$. 
We use a wide prior range on the relative magnification: $0 \leq \mu \leq 4$. 
However, if the peak width is found to be very large, more than $20$ days width at 80\% of peak height, we adjust the prior ranges on the time delay and relative magnification -- see Appendix~\ref{sec:apxcode} for details. 

\item If 2 peaks identified: the time delay is high, and we can crudely estimate it as roughly the difference in the peak positions in time, say $\delta t$. Also we have an estimation for the relative magnification,   roughly  the ratio of the two peak heights, say $r$. We then use the following priors: $(\delta t -15) \leq \dt \leq (\delta t +15)$ and $r/2 \leq \mu \leq 1.5r$.

\item If more than 2 peaks identified: we consider the two most prominent peaks and follow the treatment above. Future work will deal with greater than two image systems. 

\end{itemize} 

We follow the above steps for $g$, $r$ filters separately. If one filter has one peak while the other has two or more, we follow the two peak procedure if the separation between the identified peaks of the two peak filter is less than 30 days, otherwise we follow the one peak procedure. (The logic is that for long true time delays, we would see two peaks in both filters; this step removes the spurious fitting of a noise fluctuation far on the tail of the lightcurve as a second peak.) 
We tested this algorithm extensively as 
described in Sec.~\ref{sec:test}.

\subsubsection{Priors on hyper-parameters} \label{sec:priorhyp} 

The priors on the hyper-parameters are given 
in Table~\ref{tab:priors}. The fit generally prefers the hyper-parameters to be somewhere in the middle of the prior ranges, e.g. $m_j \sim 3.5$ and $\sigma_j \sim 0.6$. See Appendix~\ref{sec:apxcode} for further details, as well as Sec.~\ref{sec:accu} for when some of the nonphysical hyper-parameter priors may be informative. 
In practice we treat the normalization more 
subtly than the hyperparameter $N_j$, breaking 
out a scaling based on the maximum flux observed and then a zeroth order basis function $C_{0,j}$; 
see Appendix~\ref{sec:apxnorm} for details and 
Table~\ref{tab:priors} for the prior on $C_{0,j}$.

\begin{table}[tb]
\begin{center} 
\begin{tabular}{c|c} 
Hyperparameter & Prior \\ 
\hline 
$\sigma_j$ & \ \ [0.4,\,0.8] \ \ \\ 
$b_j$ & [3.0,\,4.1] \\ 
$C_{0,j}$ & [0.4,\,2] \\ 
$C_{k,j}$ & $[-5,\,5]$ \\ 
$t_1$ & $[-5,\,5]$ \\ 
\end{tabular}  
\end{center} 
\caption{Priors on the fit hyperparameters. 
Here $C_{k,j}$ is for $k=1,2,3,4$. 
} 
\label{tab:priors} 
\end{table}

In the fits 
we use STAN \cite{stan_2017,pystan} to run Hamiltonian Monte Carlo (HMC) with $7$ chains,
each 
with $6000$ iterations and $1000$ warm up steps. We check the convergence of the chains by ensuring that the \citet{Gelman_Rubin_1992} convergence diagnostic parameter $\hat{R}< 1.05$ as suggested by the STAN development team \cite{stan_2017}.

\section{Results from Testing and Validation} \label{sec:test} 

Using simulated data 
we apply the 
method proposed in Sec.~\ref{sec:method} in four ways to test the 
results obtained for time delays. 

\begin{enumerate} 
\item Systematic approach: 
We analyze systems scanning through various values for the time delays, 
relative image magnifications, and noise levels, and find in what portion 
of this parameter space the method succeeds. These lightcurves 
are generated using the Hsiao template \citep{Hsiao:2007pv}
with no microlensing,
and observing conditions matched
to those of \cite{Goldstein:2018bue}. We refer to this simulation tool based on the SNCosmo python package \cite{barbary_2014_11938} as `LCSimulator' throughout this article.


\item Blind test:  
One author used LCsimulator to generate a set of unresolved 
lensed 
SN light curves 
with a 
variety of data properties, including several unlensed systems, and provided  
only the final summed lightcurve data to another author to be fit. This 
enables robust testing, as well as assessing false positives. 

\item Applied test: 
More sophisticated lensed SN light curve simulations from 
\cite{Goldstein:2018bue} (``the Goldstein set'') 
with less 
tightly controlled noise properties are used to assess the fitting method. 

\item Validation: 
Once the method is finalized we return to untested systems from the 
Goldstein simulated data set with even less 
tightly controlled noise properties and fit the unresolved lensed lightcurves. 
\end{enumerate}

\subsection{Systematic Studies} \label{sec:3a}

\begin{figure}
\centering
\includegraphics[width=0.485\textwidth]{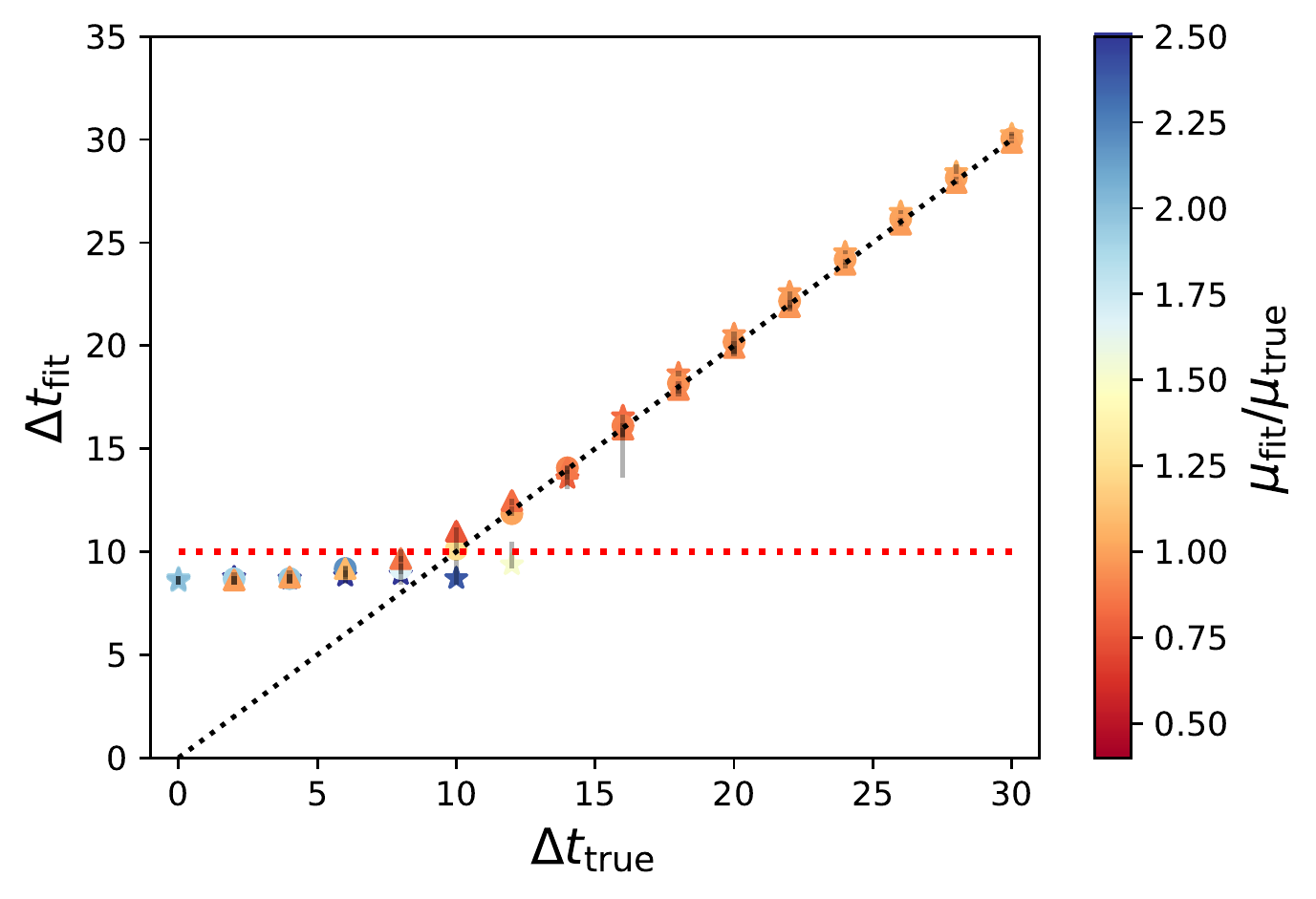} 
\includegraphics[width=0.485\textwidth]{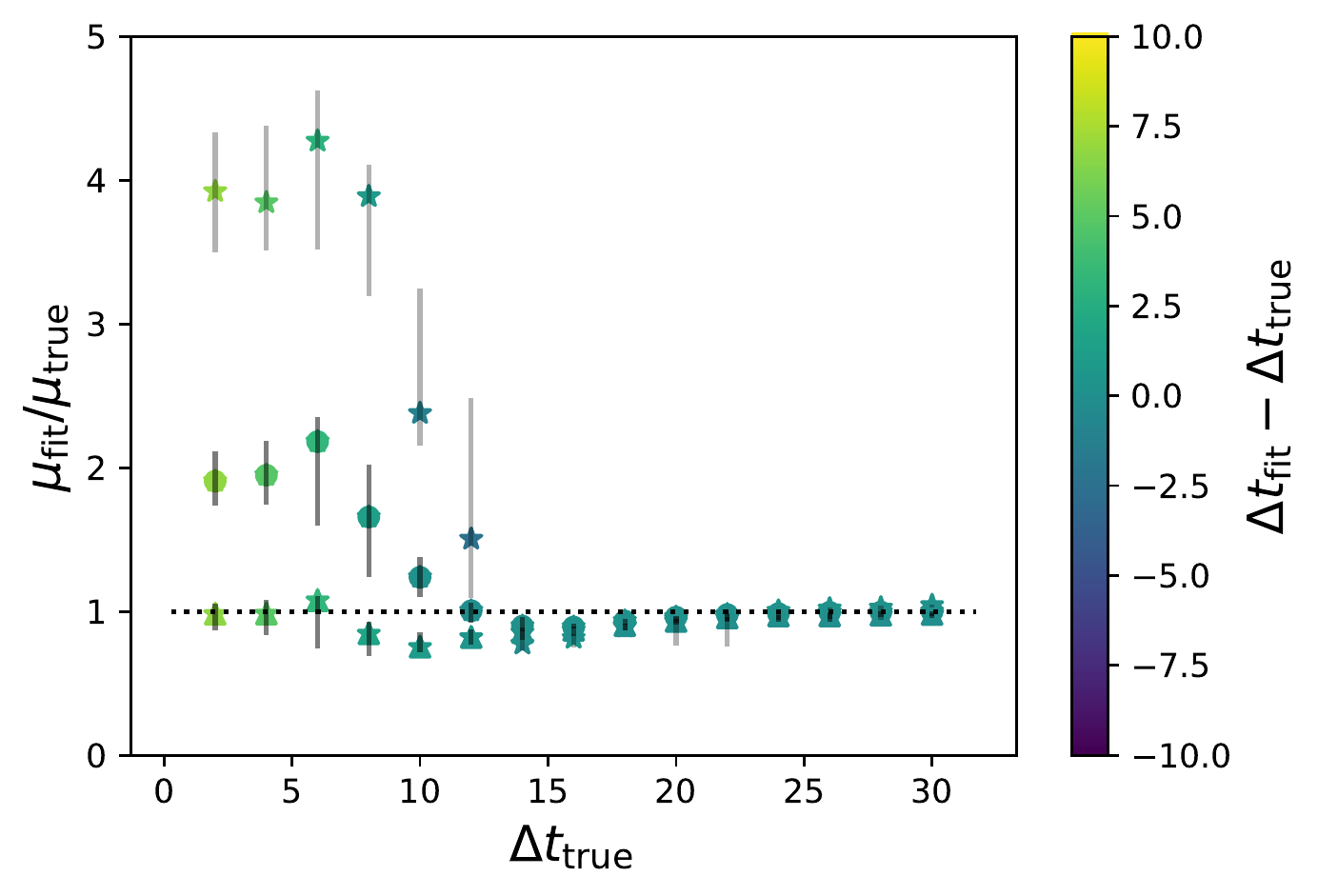} 
\caption{Noise level $=0.5\%$ (Set 1): We show $\dtbf$ (left panel) and $\mubf/\mut$ (right panel) vs $\dtt$. 
We obtain faithful results for fits giving time delays more than $\dtbf=10$ days (dashed red horizontal line in the left panel). 
Stars, circles, and triangles represent $\mut=0.5,1,2$ respectively, while color represents $\mubf/\mut$ (left panel) and $\dtbf-\dtt$  (right panel). 
The errorbars are $95\%$ quantile around the median. 
}
\label{fig:sys_check_0.5}
\end{figure}

\begin{figure}
\centering 
\includegraphics[width=0.485\textwidth]{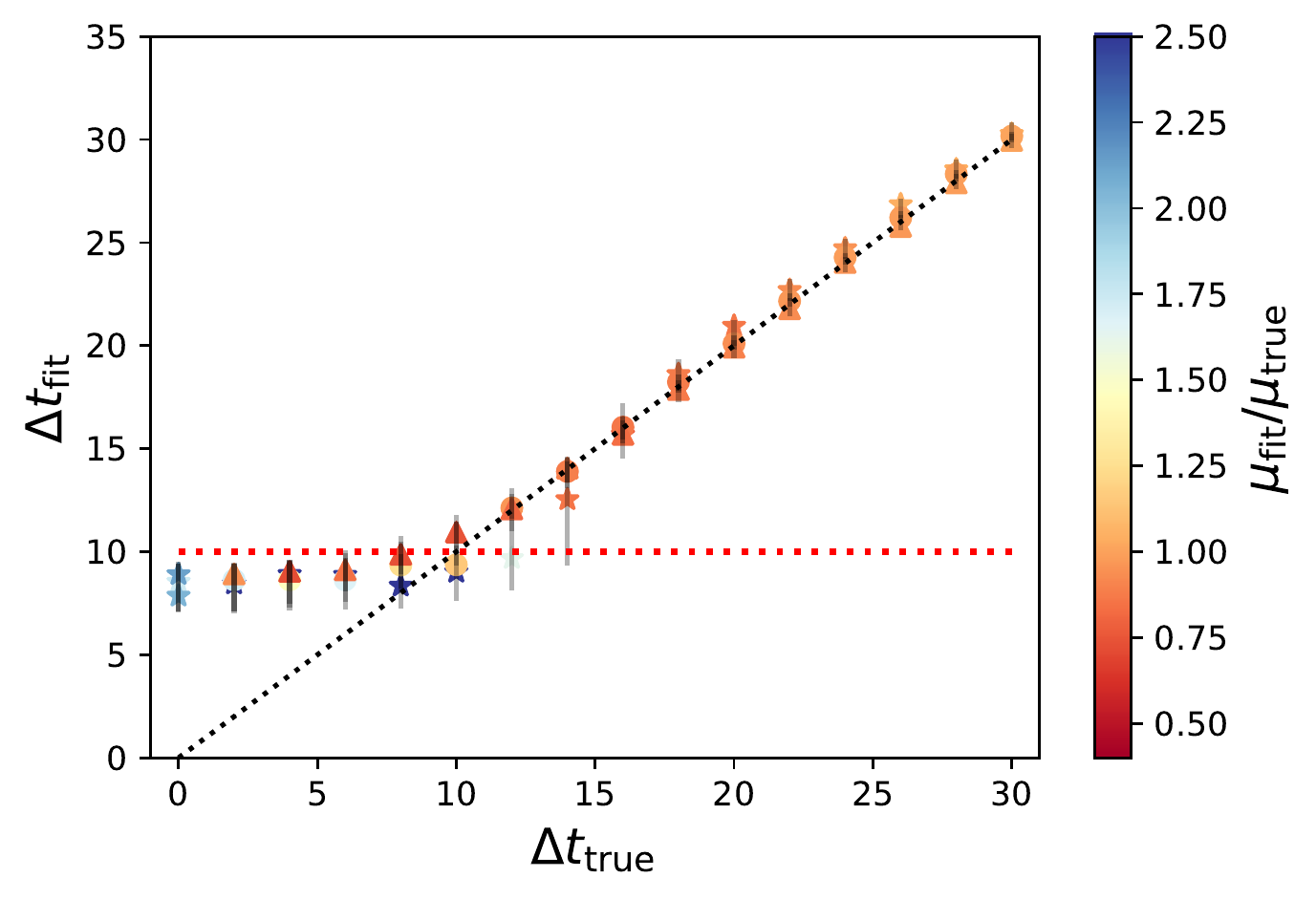} 
\includegraphics[width=0.485\textwidth]{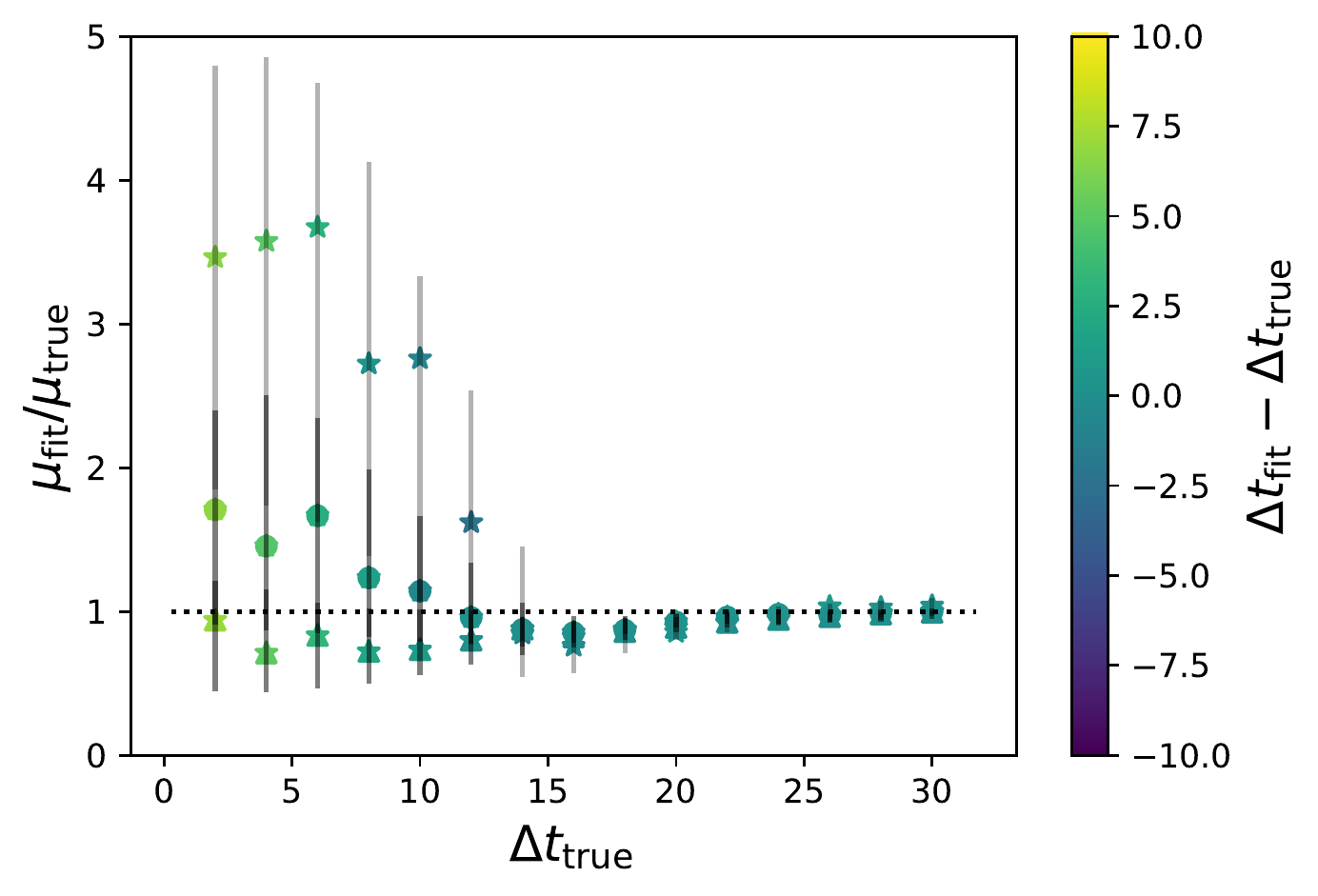} 
\caption{As Fig.~\ref{fig:sys_check_0.5}, but for noise level $=2.5\%$ (Set 2). 
}
\label{fig:sys_check_2.5}
\end{figure}

\begin{figure}
\centering 
\includegraphics[width=0.485\textwidth]{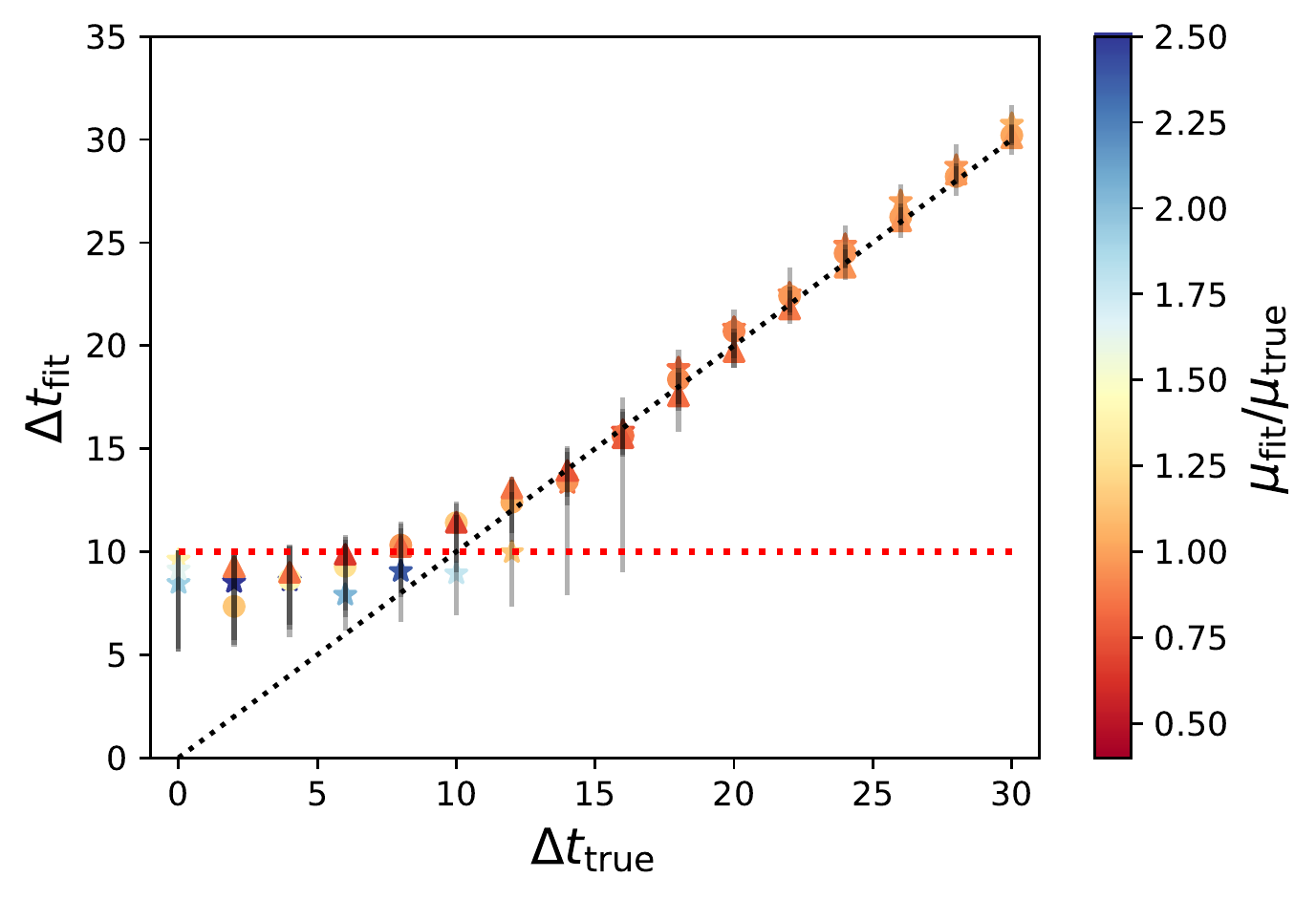} 
\includegraphics[width=0.485\textwidth]{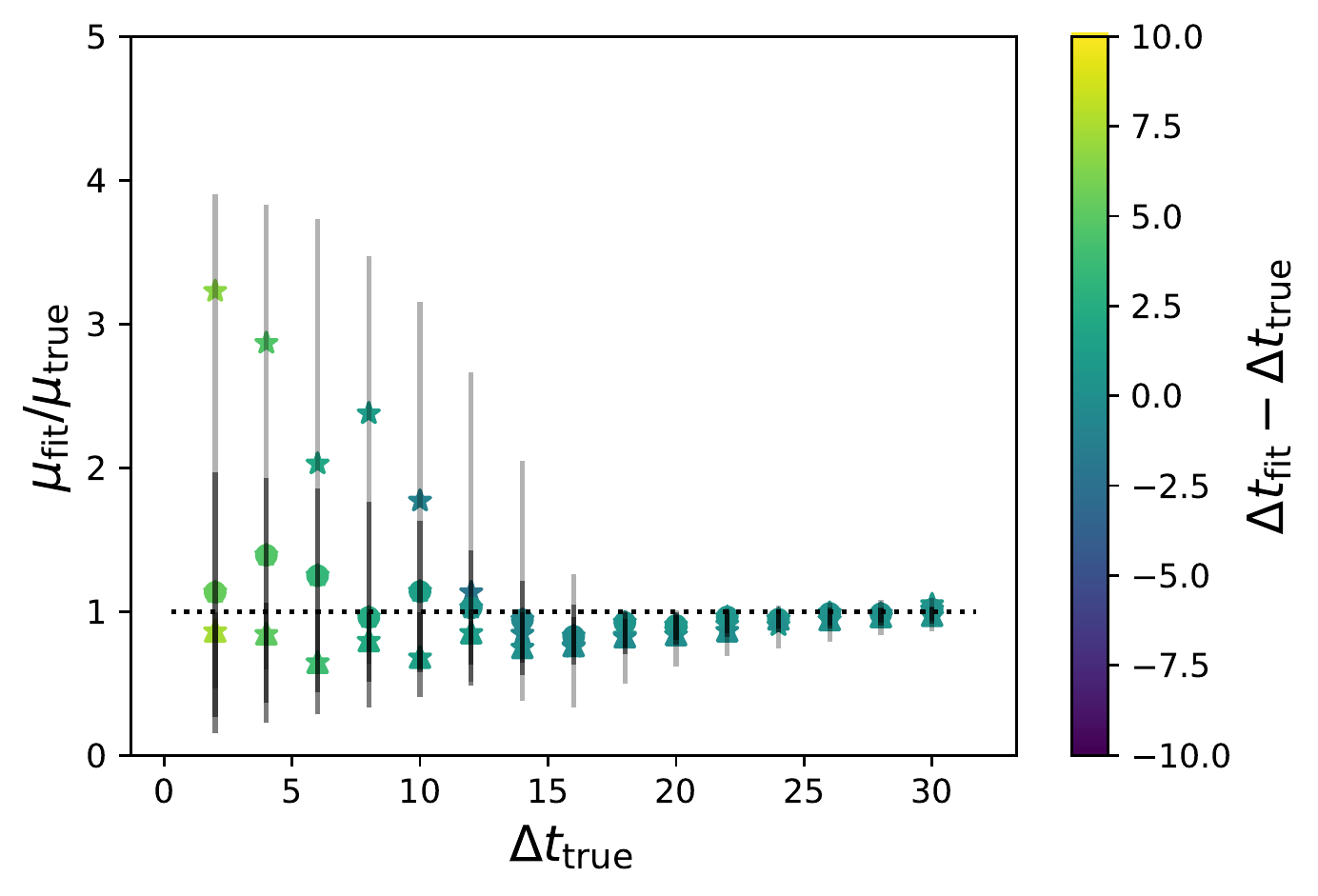} 
\caption{
As Fig.~\ref{fig:sys_check_0.5}, but for noise level $=5\%$ (Set 3). 
} 
\label{fig:sys_check_5.0}
\end{figure}

\begin{figure}[hbt]
\centering 
\includegraphics[width=0.485\textwidth]{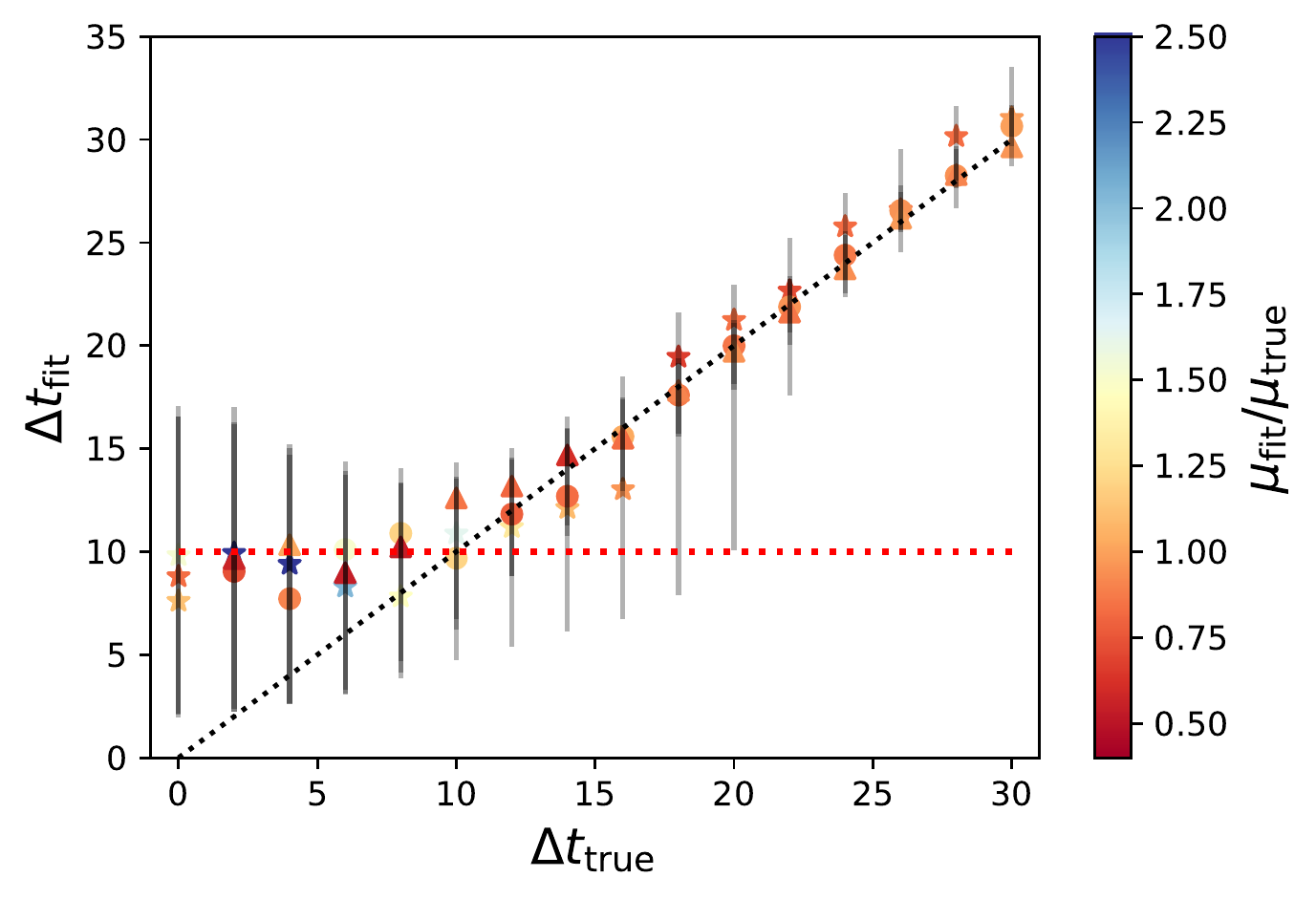} 
\includegraphics[width=0.485\textwidth]{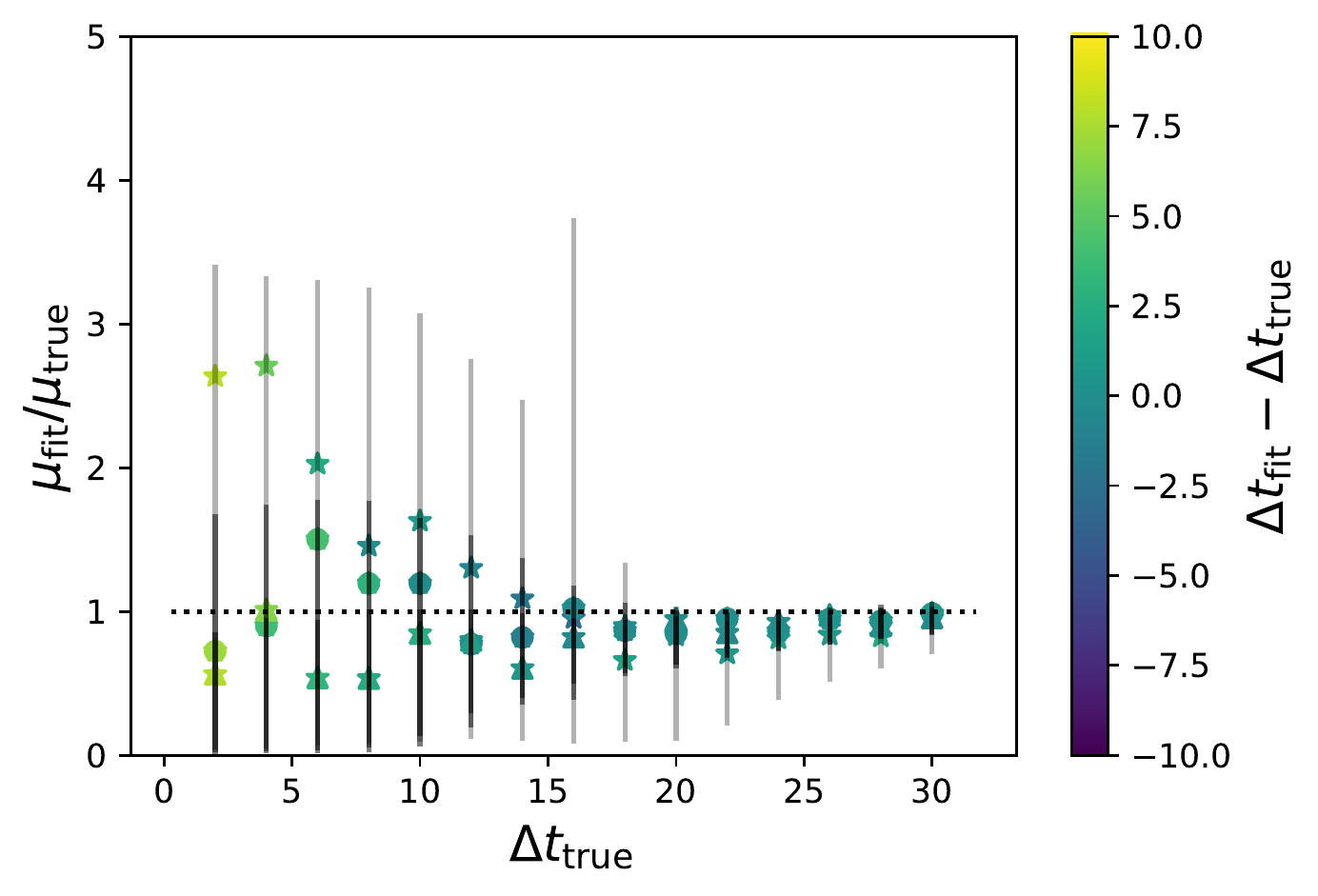} 
\caption{
As Fig.~\ref{fig:sys_check_0.5}, but for noise level $=10\%$ (Set 4). 
} 
\label{fig:sys_check_10.0}
\end{figure}

\begin{figure}[hbt]
\centering
\subfigure[$\ \mut=0.5$]{
\includegraphics[width=0.323\textwidth]{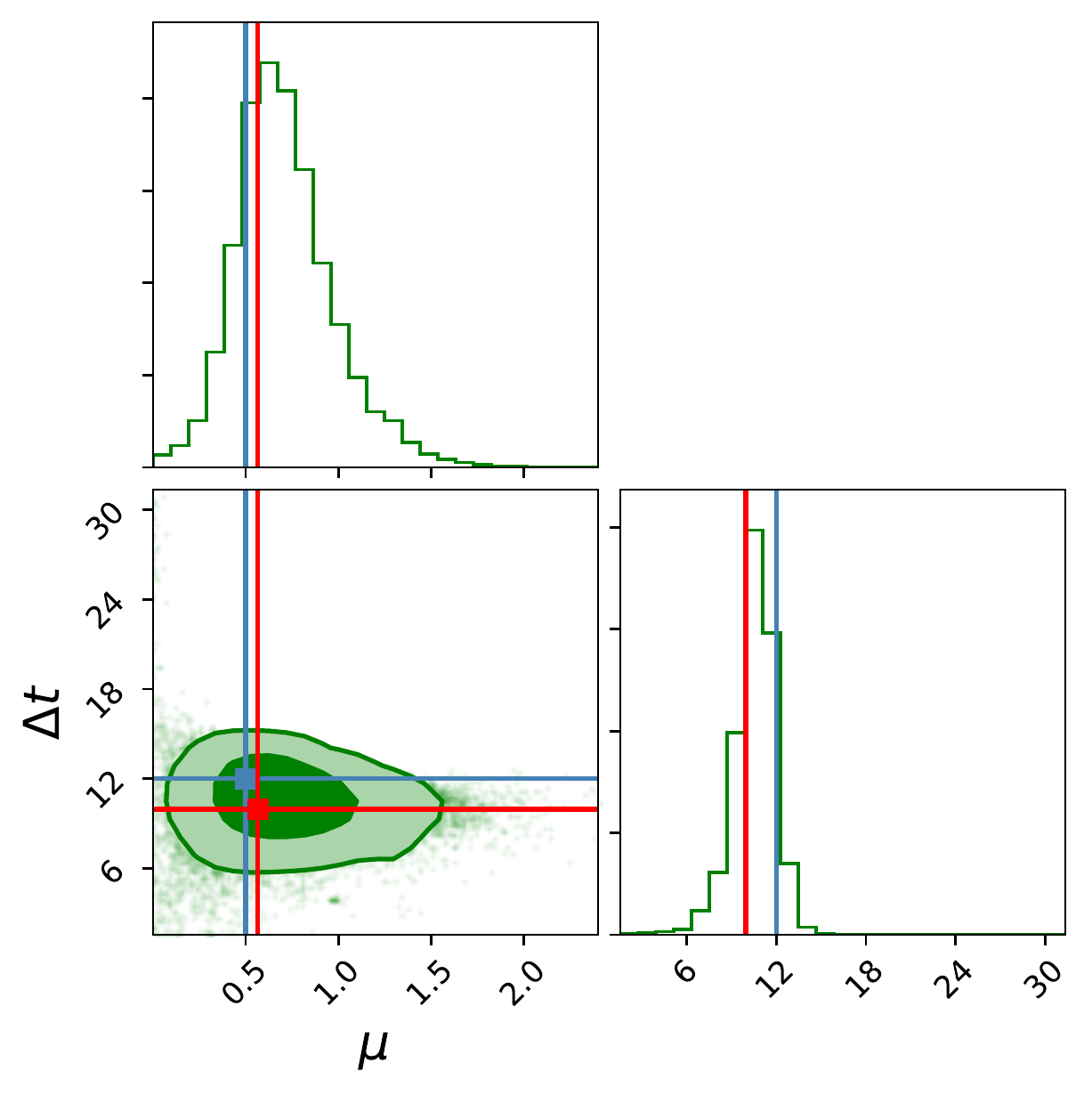}}
\subfigure[$\ \mut=1.0$]{
\includegraphics[width=0.323\textwidth]{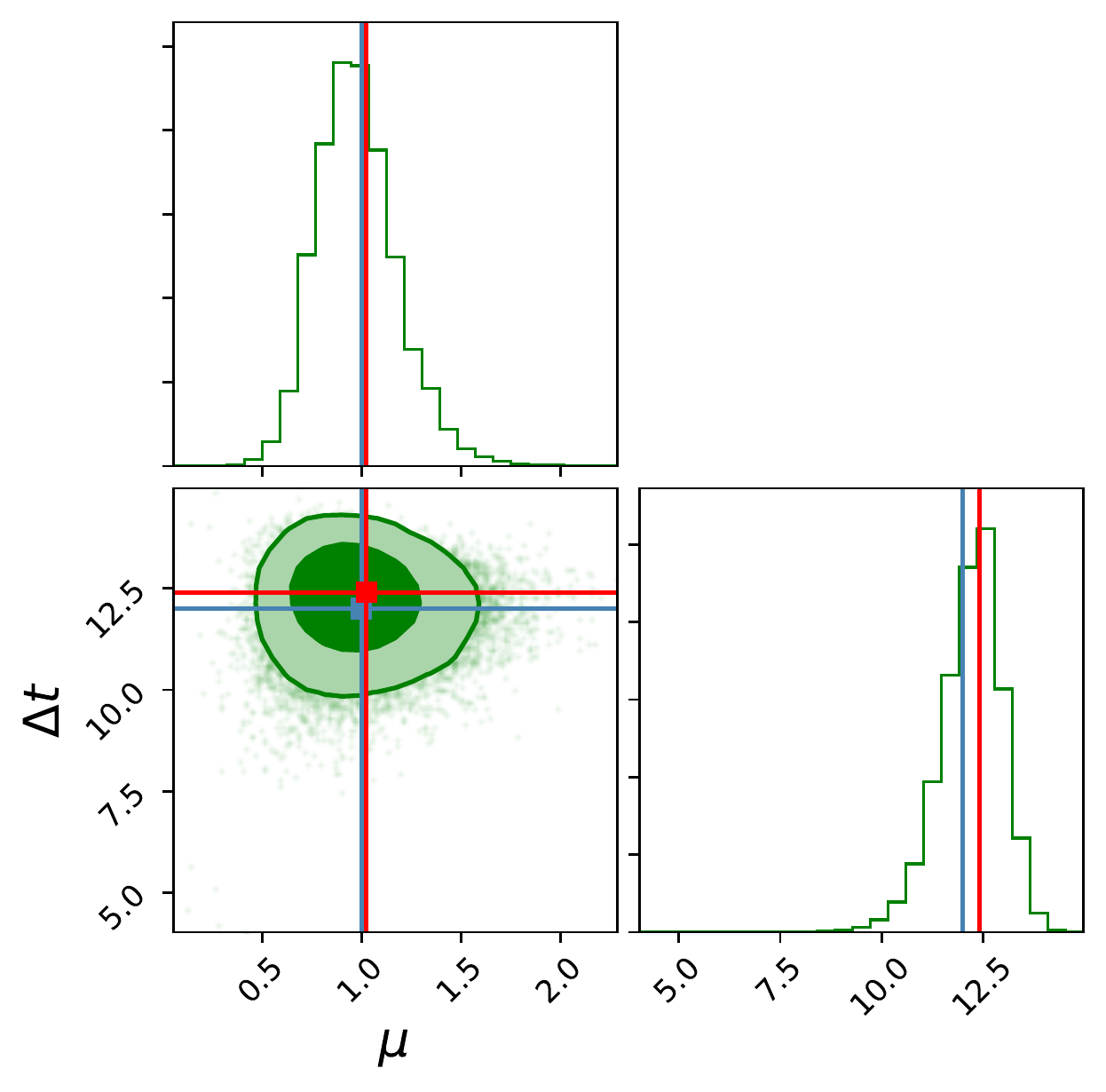}}
\subfigure[$\ \mut=2.0$]{
\includegraphics[width=0.323\textwidth]{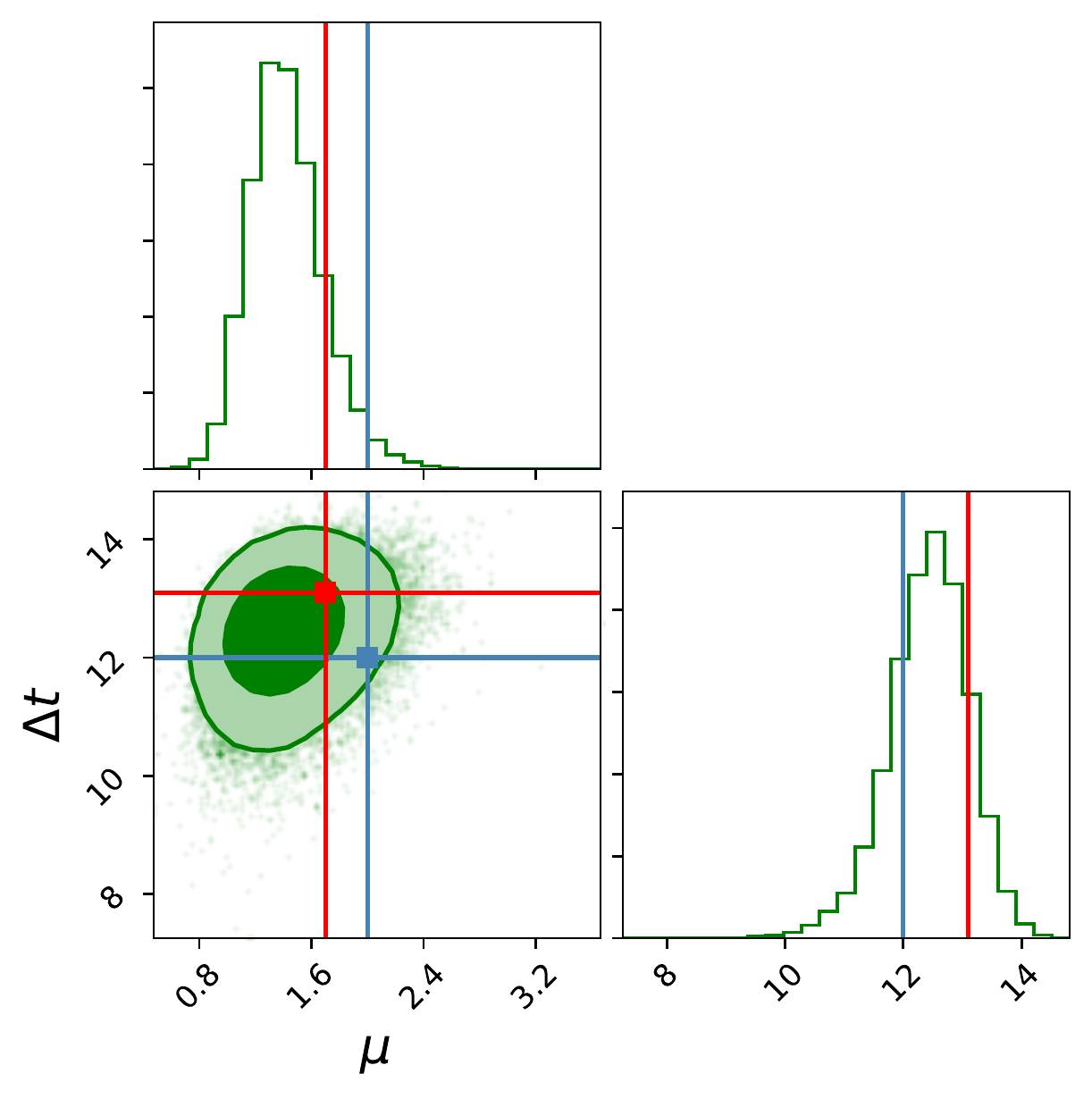}}
\caption{Corner plot showing the 2D joint probability distribution, and 1D PDFs, for $\dt$ and $\mu$, marginalizing over all other parameters. This example is for noise level $=5.0\%$ (Set 3), with  $\dtt=12$ days, and three different relative magnifications: $\mut=0.5, 1.0, 2.0$. The blue  crosshairs and points show the simulation input values while the red ones show our best-fit values. 
}
\label{fig:sys_check_5.0_dt12_cont}
\end{figure}

\begin{figure}[hbt]
\centering
\subfigure[$\ \mut=0.5$]{
\includegraphics[width=\textwidth]{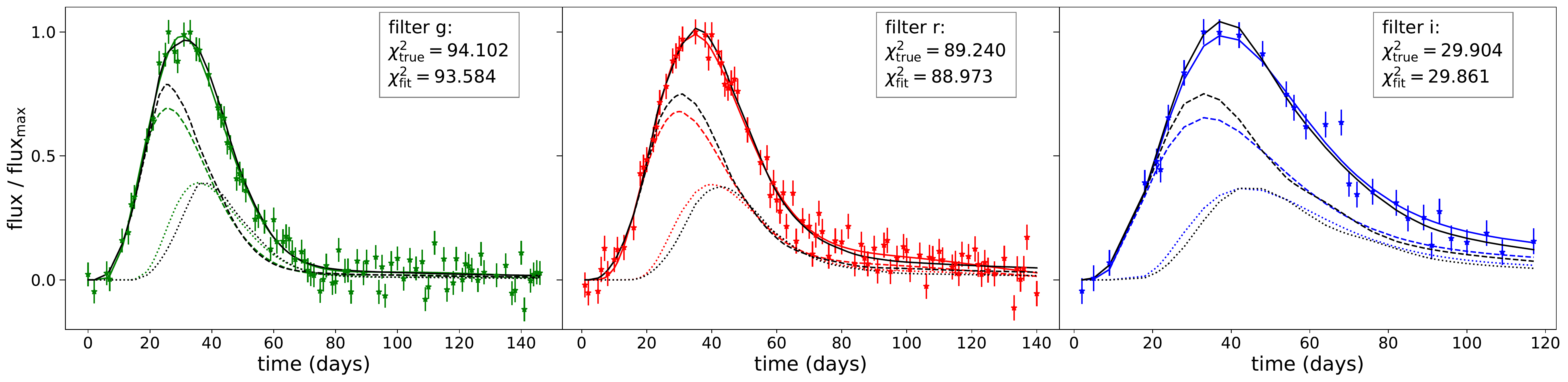}}\\
\subfigure[$\ \mut=1.0$]{
\includegraphics[width=\textwidth]{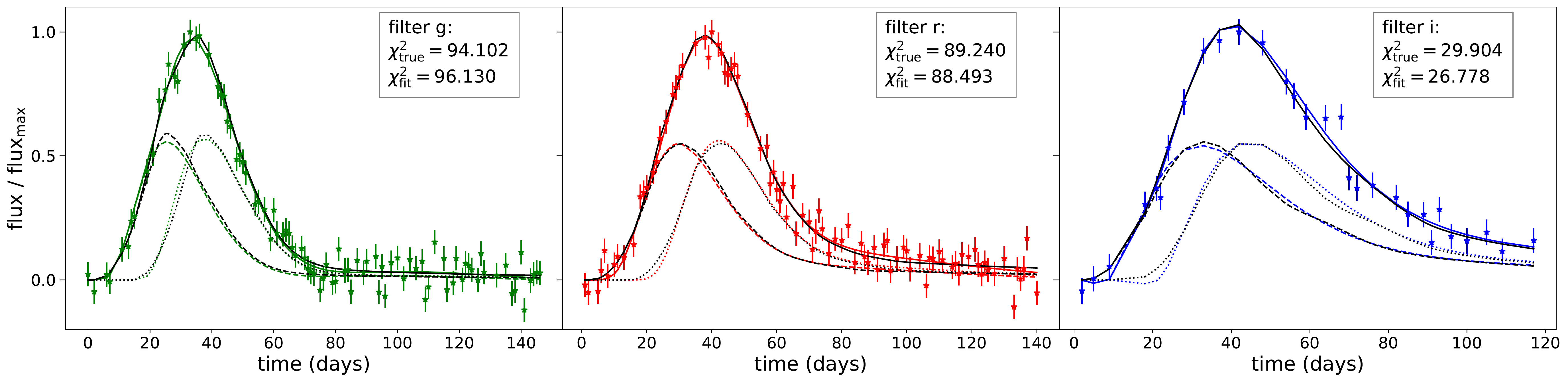}}\\
\subfigure[$\ \mut=2.0$]{
\includegraphics[width=\textwidth]{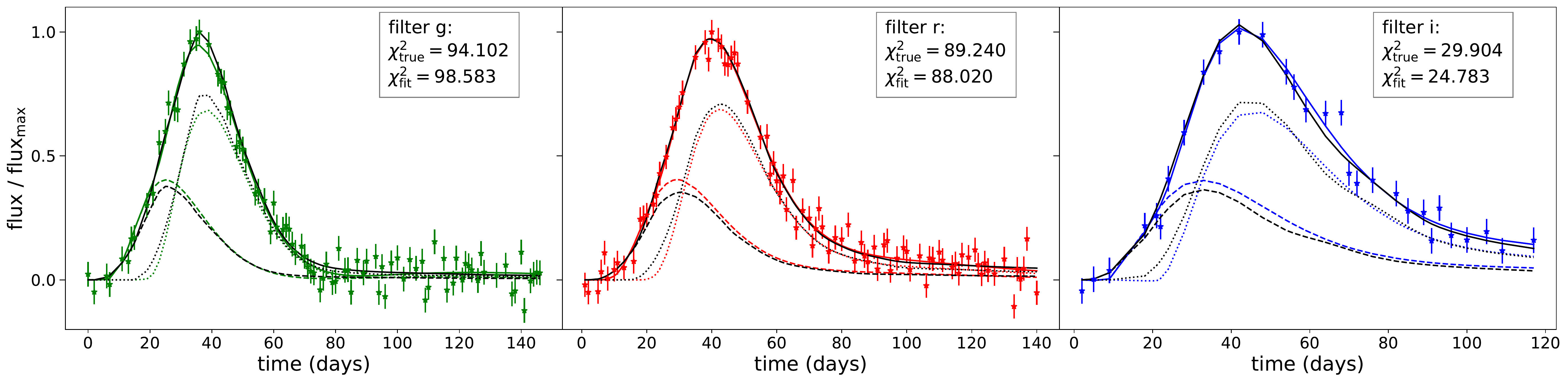}}
\caption{Simulated data with error bars (noise level $=5.0\%$, Set 3) is shown for a system with 
$\dtt=12$ days. The columns show $g$, $r$, $i$ bands and the rows show three different relative magnifications for  the images: $\mut=0.5, 1.0, 2.0$  (top, middle, bottom panels respectively). 
The black curves are the true light curves: dashed and dotted, and solid black curves represent the true light curves of the first and second images, and their combination. The colored curves are our equivalent reconstructions in each filter using 
the best fit parameters. 
While the key science quantity is merely the time delay, we do fit well the entire light curve, even for this 
moderately low time delay $\dtt = 12$ days, for all the three magnifications.
This is quantified through the $\chi^2$ values listed for the true and the reconstructed curves. 
}
\label{fig:sys_check_5.0_dt12_image}
\end{figure}

In this section we study a number of two-image systems simulated with properties scanning through time delay, relative magnification, and noise. We consider time delays $\dt=0,2,4,6,....,28,30$; i.e.\ a total of $16$ time delays. For each time delay, we simulate three
systems having $\mu=0.5,1,2$. Therefore, we have $16 \times 3 =48$ systems in a set. We then  simulate 4 such sets with  different noise levels: $0.5\%,~ 2.5\%, ~5\%$ and $10\%$ of the maximum flux. The main goal of this exercise is to test how our method performs in different conditions, finding the range of validity in terms of the time delay, magnification, and noise level (or alternately, ``testing to destruction''). Set 1 has an  extremely small noise level, quite unrealistic, and serves to establish fundamental limits in the method. Set 4 has an extremely large noise level, where we expect large uncertainty in our estimations.  Note that each set has $48$ systems out of which $3$ are truly unlensed with $\dt=0$ (and several others have time delays less than some surveys' cadence). 

The summary of the results are shown in Figures~\ref{fig:sys_check_0.5}, \ref{fig:sys_check_2.5}, \ref{fig:sys_check_5.0}, \ref{fig:sys_check_10.0} for the 4 sets (each with a different noise level).
Each plot has 48 systems. The three markers, `star', `filled circle', `triangle', represent $\mut=0.5,1,2$ respectively. The errorbars shown here (and throughout this paper) are 
discussed in Sec.~\ref{sec:accu}. 
The colourbars represent the quantities $(\mubf/\mut)$ in the left panels and $\dtbf-\dt_{\rm true}$ in the right panels. Note that for unlensed 
systems we have both $\dtt=0$, $\mut=0$; so we set
arbitrarily 
the colour according to $\mubf$ for such systems and mark them by stars in the left panels; all systems with $\mubf/\mut>2.5$ (basically bad fits for $\mu$) have been marked with the dark blue colour for $\mubf/\mut=2.5$. The black dotted diagonals in the left panels of the figures represent $\dtbf=\dtt$ while the horizontal red dotted line is $\dtbf=10$ days below which the solutions depart from this relation. We quantify 
this further in Sec.~\ref{sec:accu}. 

From these plots we observe the following: 
\begin{itemize}
     \item For all four sets, we recover the true solutions well when $\dtt>10$ days. While the cosmology focus is on $\dtbf$, we can also find $\mubf$ reasonably well. This is fairly robust with noise level, although the 10\% level is noticeably more uncertain. 
     Figure~\ref{fig:sys_check_5.0_dt12_cont} shows an example of the 2D joint posterior, and 1D PDFs, for $\dt$ and $\mu$. 
     
     \item Furthermore, although again it is not the focus for cosmology, the individual image light curves match  well with the true ones.  Figure~\ref{fig:sys_check_5.0_dt12_image} presents an example of the image lightcurves, as well as the unresolved total lightcurve, compared to the inputs and data. 
 
 \item For $\dtt<10$ days, in all four sets $\dtbf$ does not trace $\dtt$. Instead of following the true time delay, the estimated time delay remains stable around $\dtbf \lesssim 10$ days for $\dtt<10$ days.  
This holds as well for a completely different intrinsic lightcurve approach not discussed in this paper, and seems likely to be due to the data sampling cadence. 

\end{itemize}

The bottom line is that for unresolved lensed SN with $\dtt \geq 10$ days our method can  1) identify them as lensed, 2) fit the time delay used for cosmology, and 3) further fit the magnification and individual lightcurve shapes, not essential for cosmology. When the fit points to $\dtbf<10$ days, we do not have confidence from this approach that the SN is actually lensed or its exact time delay. 

\subsection{Systematic Studies for Unlensed Systems}  \label{sec:unlens} 

In the systematic study presented in the previous section, we studied a total of $12$ unlensed systems (having $\dtt=0$ and/or $\mut=0$), 3 cases in each of the 4 sets. In all the $12$ unlensed cases, our fit estimates $\dtbf < 10$ days and hence we would  
not claim detection of a lensed SN. 

To investigate further, we study next a large number of unlensed systems, again simulated using LCsimulator. We consider 360 unlensed SN, 120 in each of 3 sets, with noise levels at $2.5\%$, $5 \%$, $10 \%$ (of the maximum of the light curve). 
If we use our acceptance criterion $\dtbf>10$ days suggested by 
Figs.~\ref{fig:sys_check_2.5}--\ref{fig:sys_check_10.0} then we have no false positives 
for the 2.5\% noise level and about 8\% 
false positives for the 5\% and 10\% noise 
levels. However, raising the acceptance to 
$\dtbf>12$ days completely eliminates false 
positives: none out of the 360 would be 
mistakenly interpreted as unresolved lensed 
SN.

\subsection{Blind Testing} \label{sec:blind} 
\begin{figure}
\centering
\includegraphics[width=0.485\textwidth]{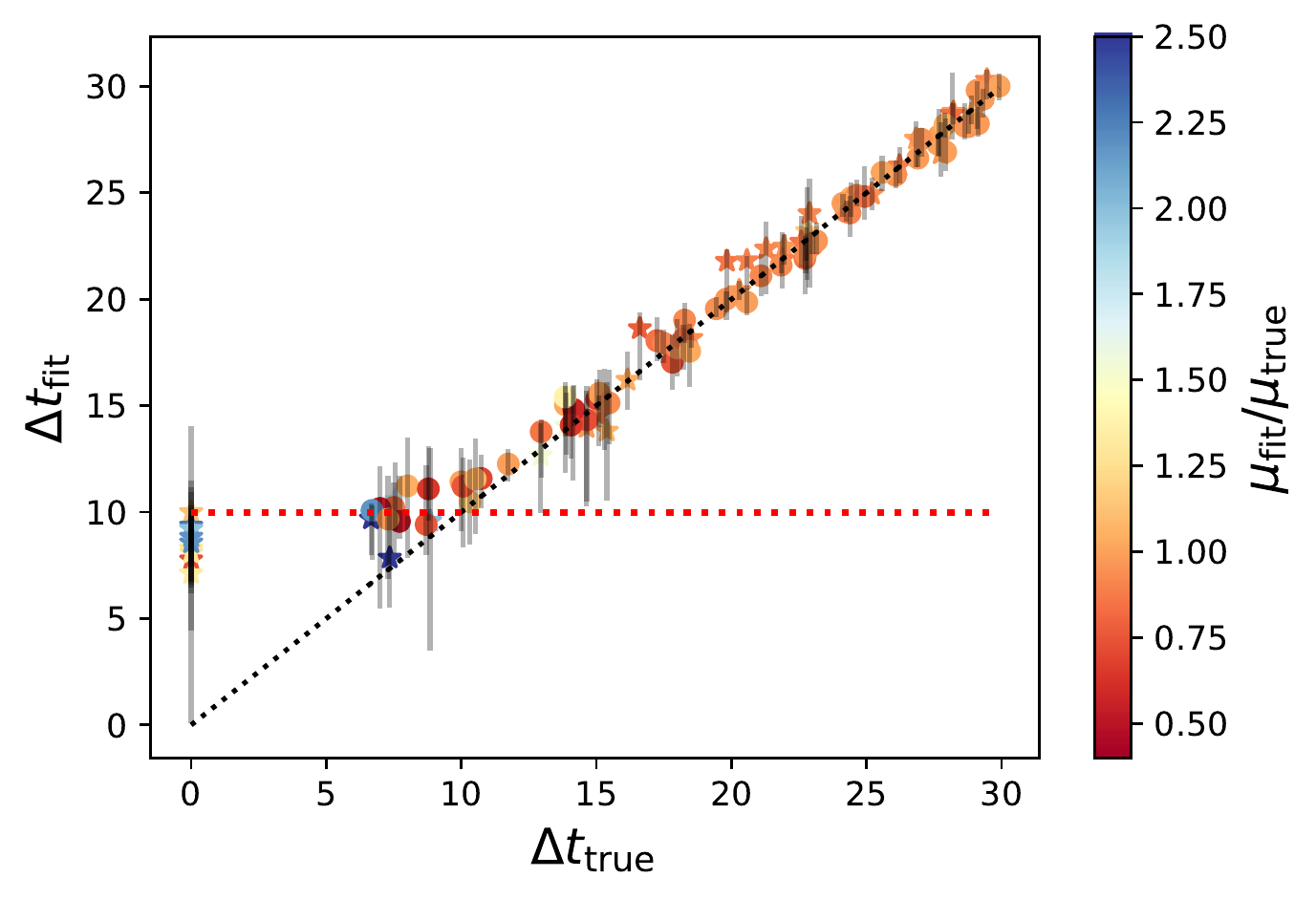}
\includegraphics[width=0.485\textwidth]{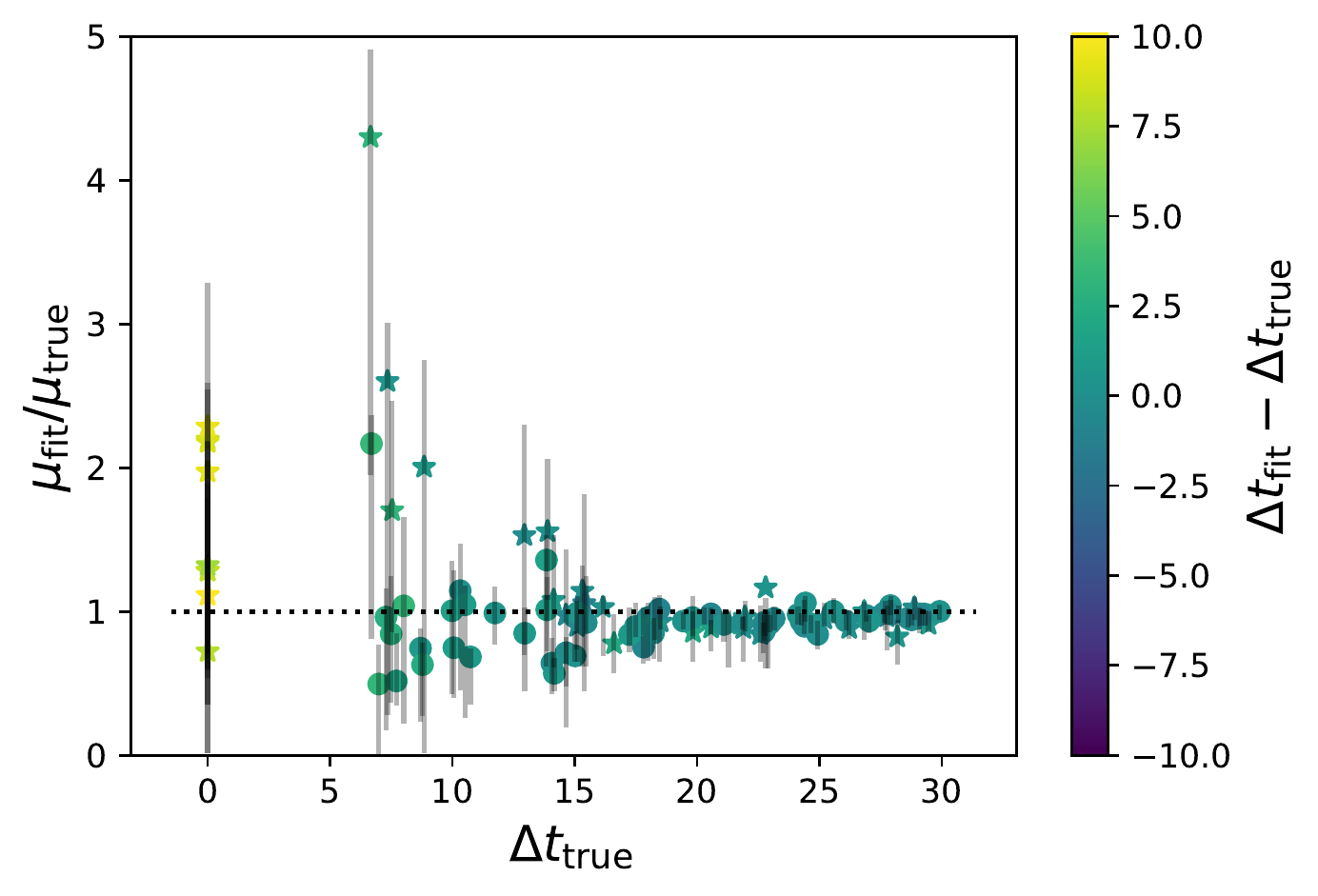}
\caption{Blind test: 
As Fig.~\ref{fig:sys_check_0.5}, but for a blind set having $100$ systems, 9 of them are unlensed while the time delays of the rest are distributed over $6-30$ days. Stars and circles represent systems with $\mut \leq 1$ and $\mut > 1$ respectively. 
} 
\label{fig:Alex_blind_set4}
\end{figure}


To test the robustness of the developed analysis pipeline, 
one author simulated a set 
of unresolved
SN light curves based on the
Hsiao template \citep{Hsiao:2007pv}, 
with no microlensing, and
observing conditions matched
to those of \cite{Goldstein:2018bue}. This set included a range of 
time delays, and also unlensed cases. Another author 
received only the final summed lightcurve data to carry out a 
fit, without further communication. 
This enabled testing without any case by case tweaking using 
knowledge that would be not  be available for real data; in  
addition it served as a further test of false positives. 

This test used $100$ systems with the noise level randomly set between $2.5\%$ and $10\%$ (of the maxima of the observed light curve). After the analysis was complete and fits frozen, it was 
revealed that $9$
systems were unlensed and the remainder had time delays  
distributed over $6$--$30$ days. 
Results are shown in Fig.~\ref{fig:Alex_blind_set4} 
and appear well in line with previous tests -- 
good tracing of the true time delay and no false 
negatives.

\subsection{Applied Tests on the Goldstein Set} \label{sec:applied} 

\begin{figure}
\centering 
\includegraphics[width=0.485\textwidth]{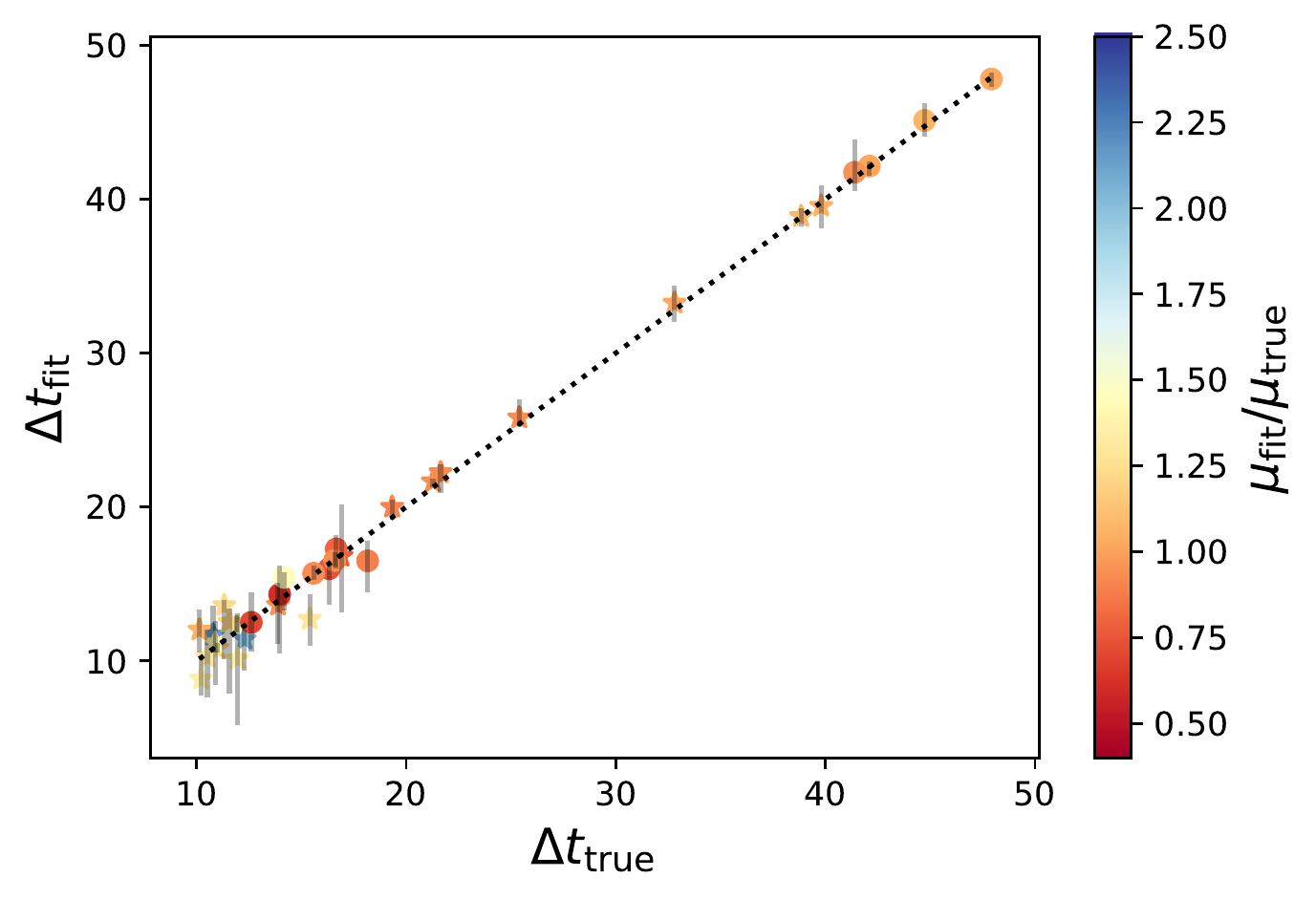}
\includegraphics[width=0.485\textwidth]{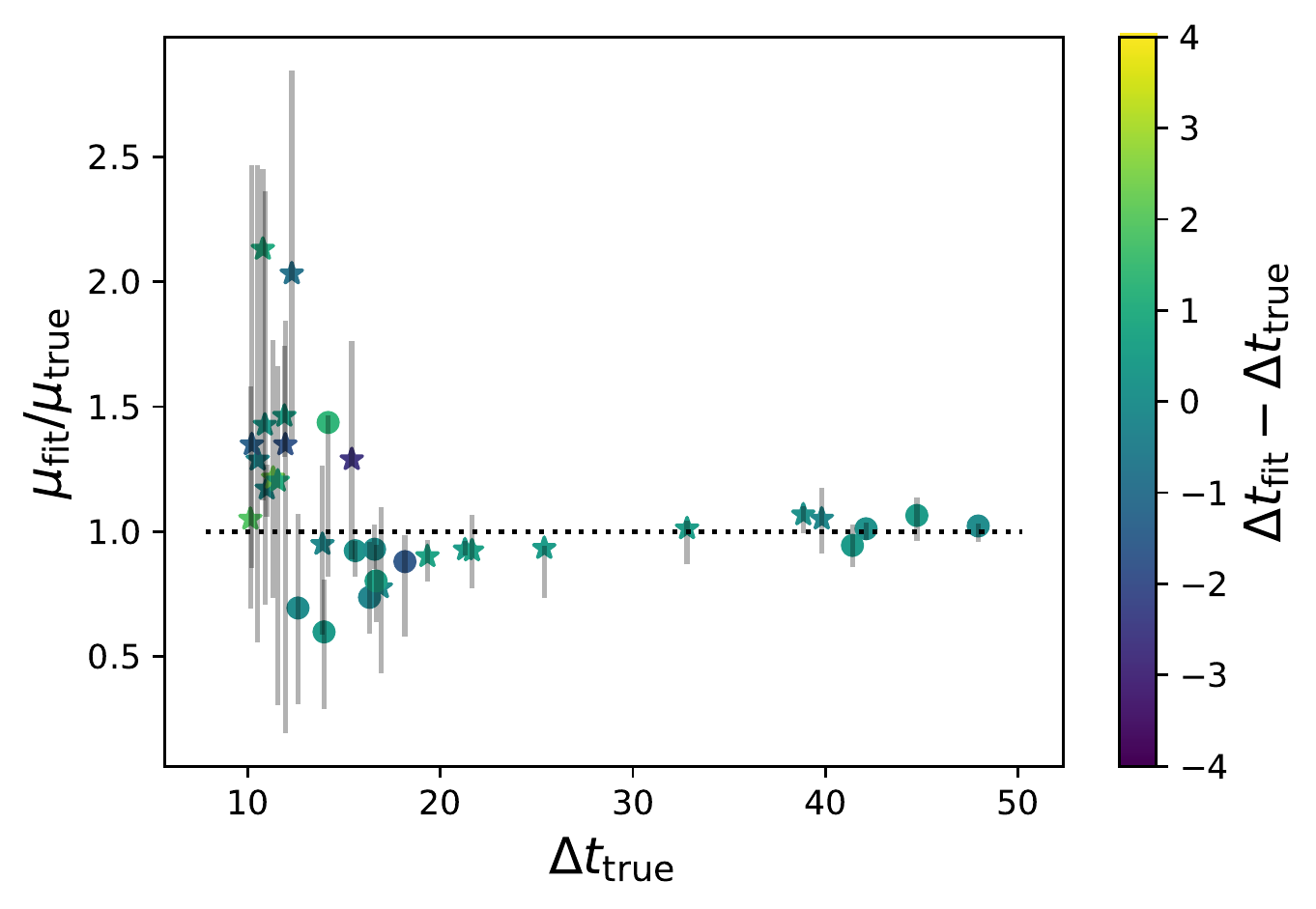}
\caption{
The best-fit time delays for 33 ZTF-like systems from \citet{Goldstein:2018bue}
are shown with respect to the true values. The colourbars show $\mubf/\mut$ and $\dtbf-\dtt$ in the left and the right panel respectively. Stars and filled circles represent systems with $\mut \leq 1$ and $\mut > 1$ respectively. The figure shows that we get excellent fits and smaller errorbars for higher time delay systems. 
}
\label{fig:bsf_scatter_error1}
\end{figure}

\begin{figure}
\centering
\subfigure{
\includegraphics[width=0.4835\textwidth]{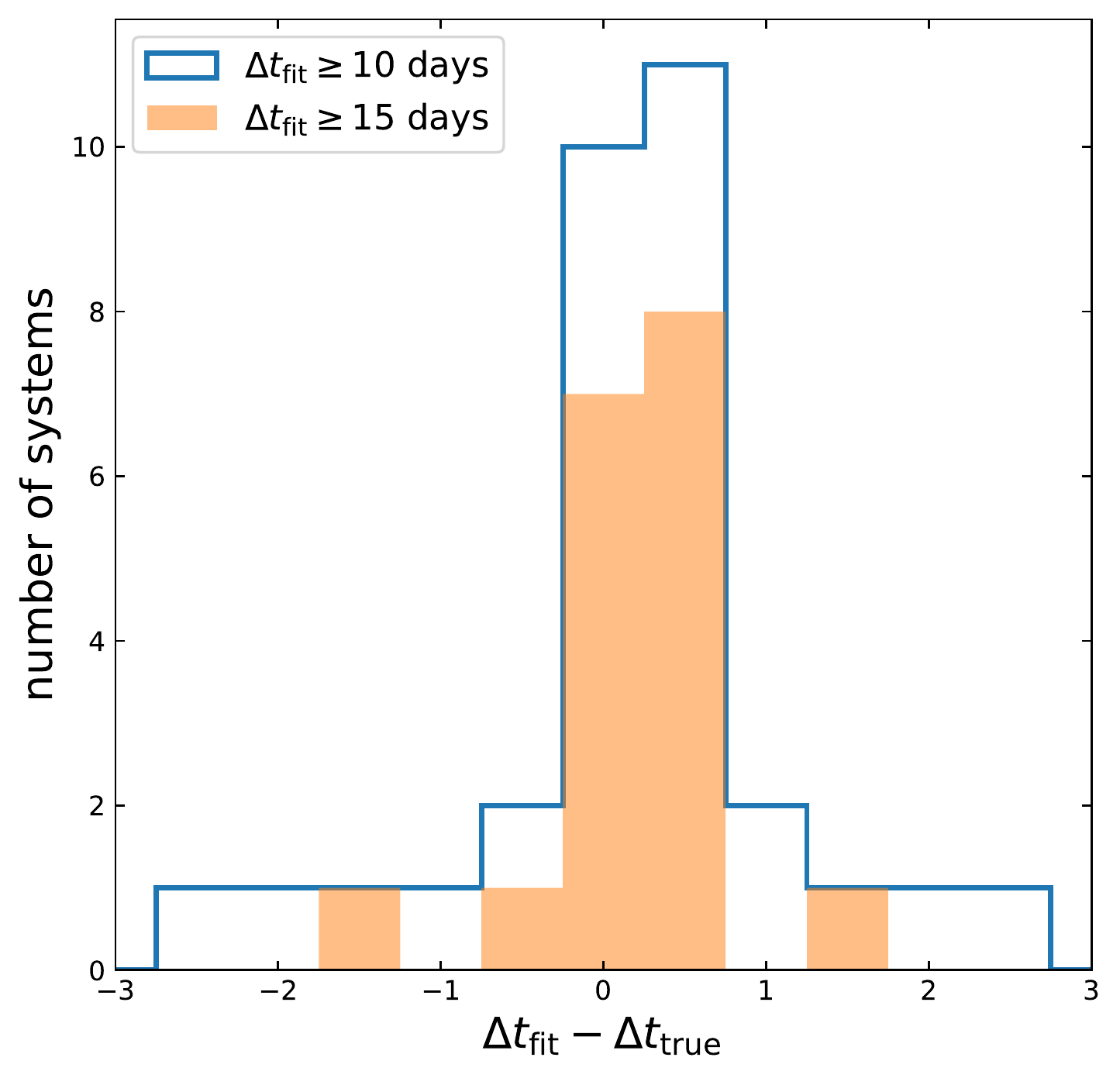}
}
\subfigure{
\includegraphics[width=0.4868\textwidth]{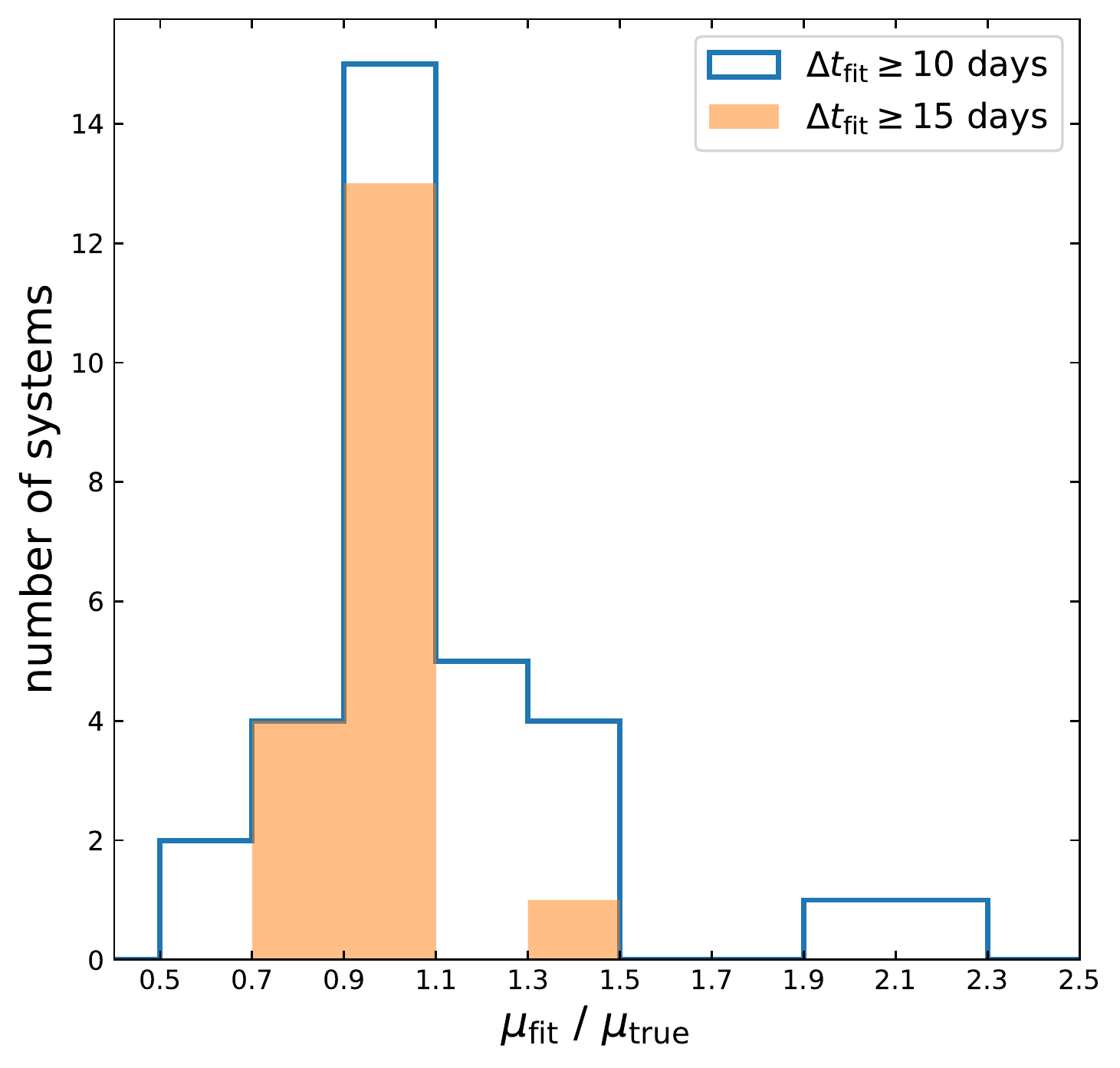}
}
\caption{The histograms of the best-fit values of time delay 
and magnification from the set of 33 ZTF-like systems are shown relative to the true values, in bins of width 0.5  days in $\Delta t$ and 0.2 in $\mu$. The blue histogram in either panel consists of systems which we could fit with confidence, i.e.\ for which $\dtbf \geq 10$ ($32$ systems), with the orange shaded regions restricted to those systems with  $\dtbf \geq 15$ days ($18$ systems). 
Note only two out of the $33$ systems in this set have $|\dtbf-\dtt|>2$ days and none have $|\dtbf-\dtt|>3$. 
We see that for most systems the fits are quite close to the true value. 
}
\label{fig:bsf}
\end{figure}

\begin{figure}
\centering 
\includegraphics[width=0.4835\textwidth]{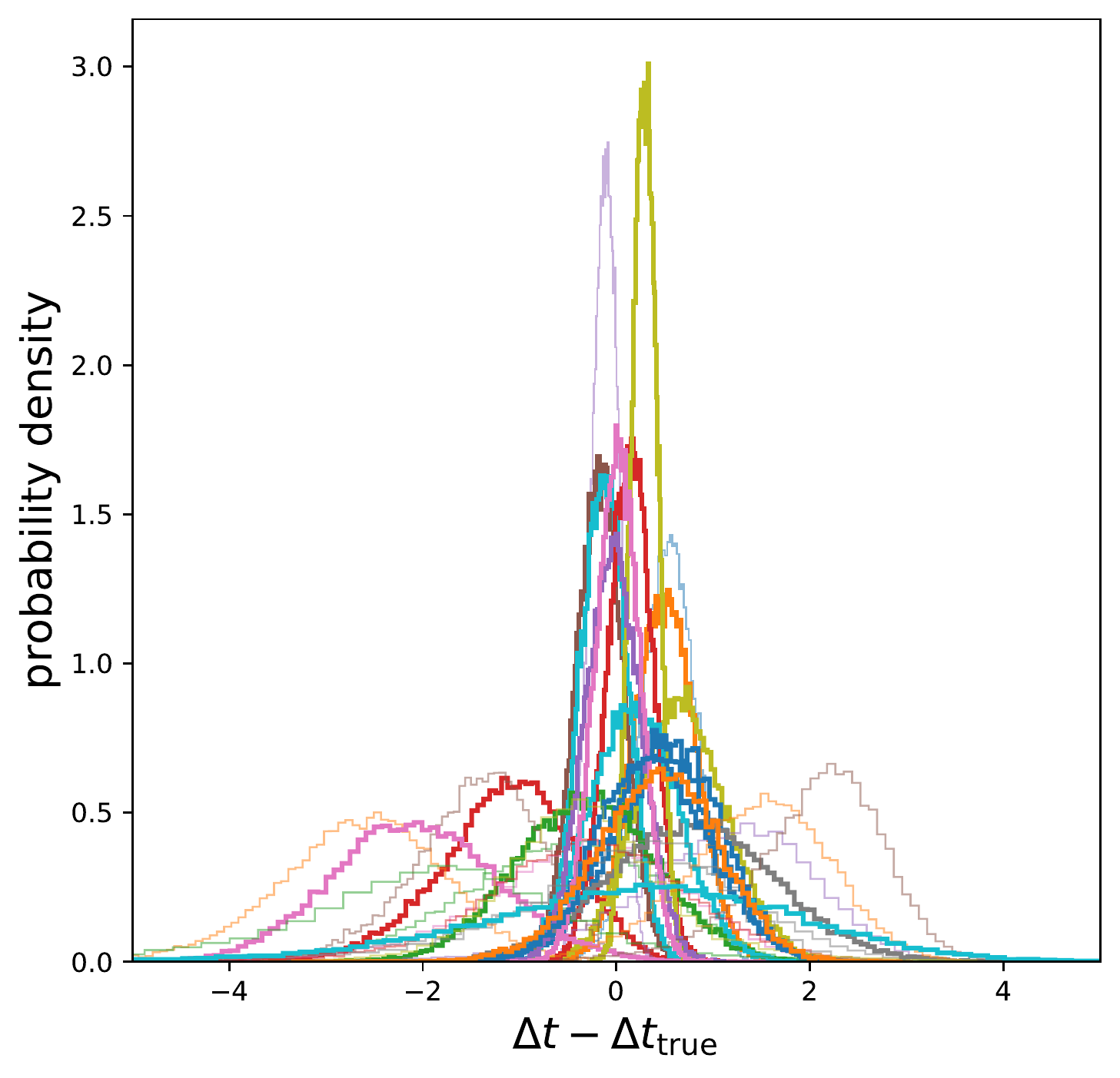} 
\includegraphics[width=0.487\textwidth]{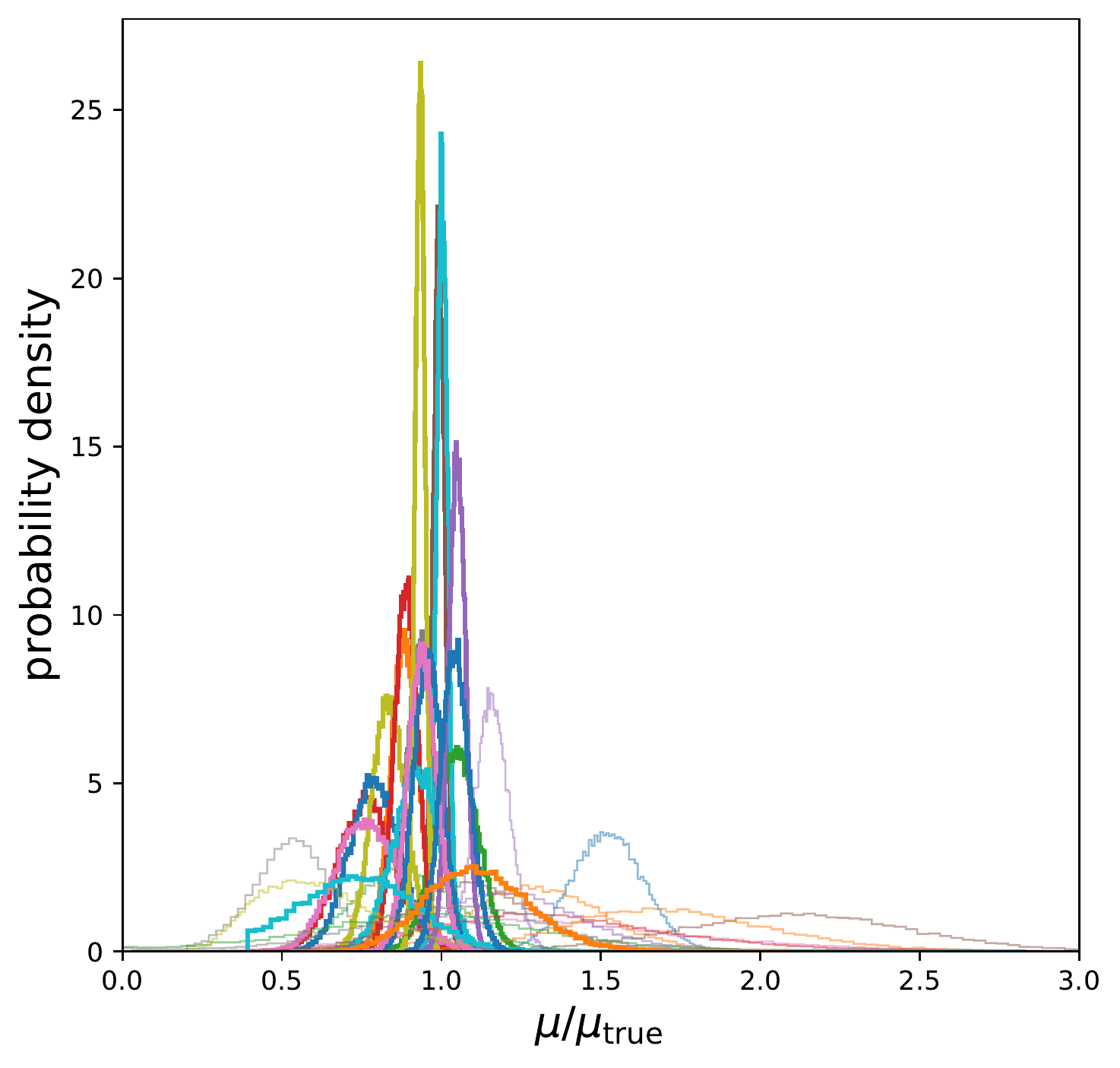} 
\caption{Rather than merely the best fit values, here we show the full 1D probability densities (PDF) of time delay 
and magnification (relative to the true values) of the 33 systems, plotted simultaneously. Again the full 1D histograms trace the true values quite well. 
Bolder curves denote systems with $\dtbf>15$ days.
}
\label{fig:bsf_overlap}
\end{figure}

For more sophisticated lensed SN simulations we employ 
the ZTF-Ia lensed SN simulations compiled in \cite{Goldstein:2018bue} that fulfill the following criteria: 1) 
Light curve visually appears smooth (and the shape is `like a light curve'); 2) The system has a minimum of $200$ data points when combining $g$, $r$, $i$ bands (typically the $i$-band data points are much fewer than for the $g$, $r$ bands) and $0.5 \leq \mu \equiv \mu_2/\mu_1 \leq 2$, $\dtt \geq 10$ days. 
This gives rise to an initial set of 33 systems, with an average noise level of $\sim7\%$.  

Our results for the time delays and magnifications appear in 
Fig.~\ref{fig:bsf_scatter_error1}. Again, the fits trace the true 
values. Note 7 of the 33 systems are at higher time delays than we 
have previously considered and the fit continues to work well 
(unsurprisingly since the time delays are higher, but still useful 
to check). Figure~\ref{fig:bsf} compiles the statistics of the fits 
by presenting the histograms of scatter of the best fit from the true values, for 
the outcomes where we accept the fit results -- either $\dtbf>10$ 
or $15$ days. The histograms are highly peaked near the true values, 
only $2$ of $33$ cases are off by more than 2 days in the time delay, and 
none by more than 3 days. 

Since more information is present in the full fit distributions, not 
just the best fit values, Fig.~\ref{fig:bsf_overlap} presents the full 
1D PDFs for the time delays and magnifications. They appear fairly  
Gaussian and the ensemble is well peaked near the true values.

\subsection{Validation Tests on the Goldstein Set} \label{sec:valid} 

Finally, following the previous full program of testing we apply the 
method to a larger sample from \citet{Goldstein:2018bue}. We consider 
systems with relaxed constraints on the relative magnification: 
$1/3 \leq \mut \leq 3$ while keeping the time delay range the 
same as the training set, $10 \leq \dtt \leq 50$ days. The other 
criteria pertaining to the data quality are kept the same except 
that for the purpose of validation, we extend the testing by 
including noisier systems. 
The set of $71$ systems 
so identified has a larger overall noise level ($\sim 10.4 \%$) 
compared to the previous samples so together with the relaxed 
constraints we therefore expect larger uncertainties  and scatter. 

\begin{figure}
\centering
\includegraphics[width=0.485\textwidth]{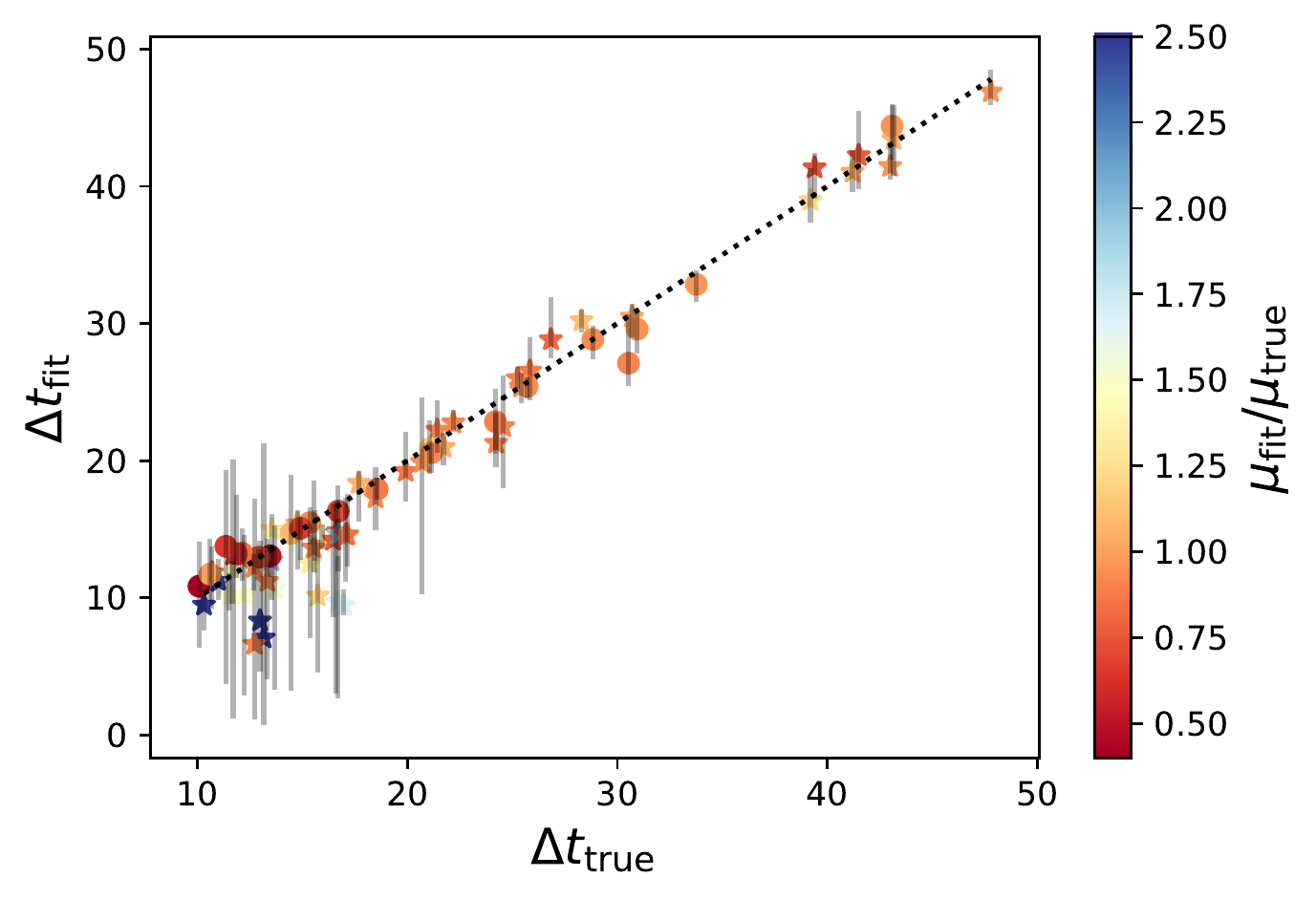}
\includegraphics[width=0.485\textwidth]{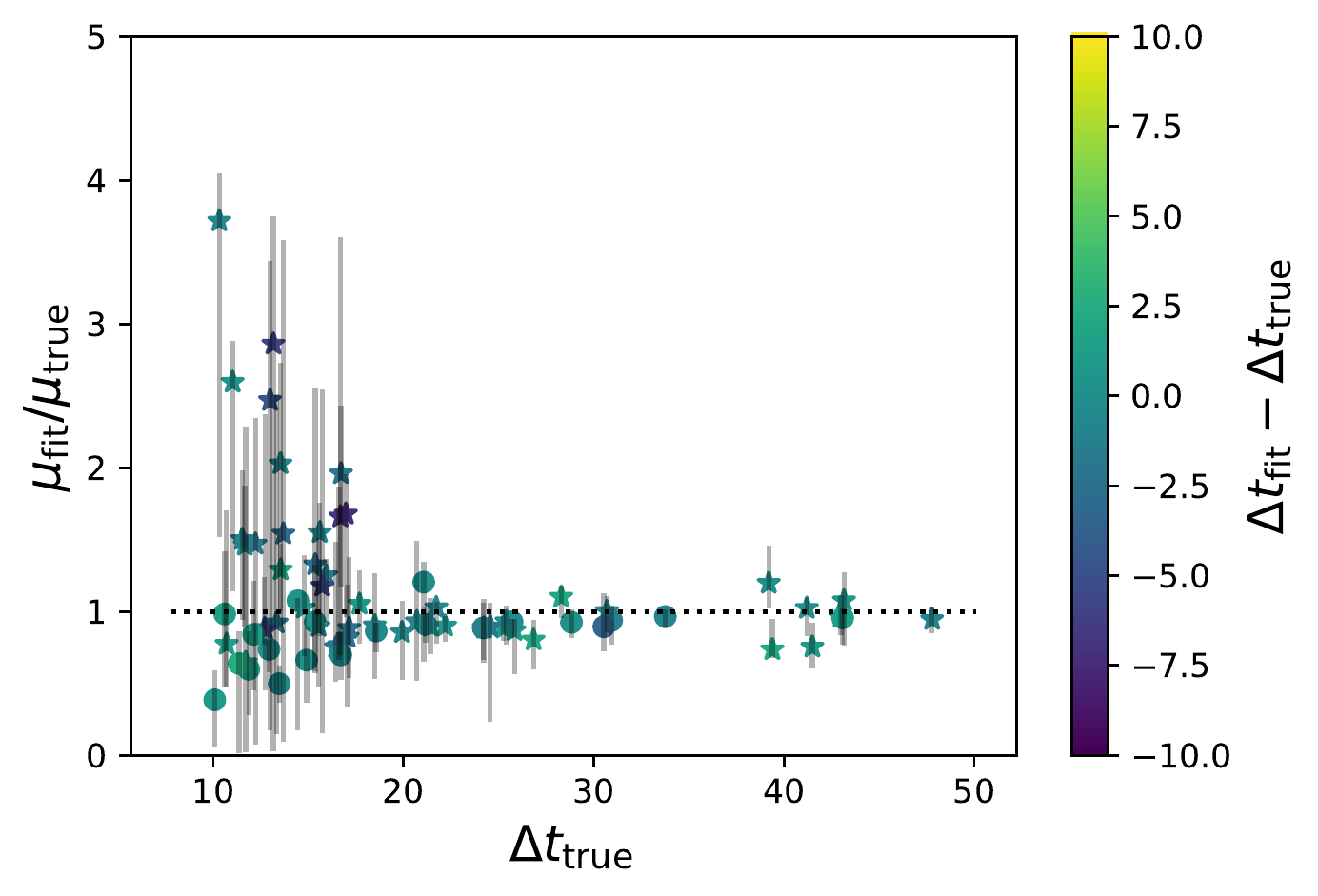}
\caption{
As Fig.~\ref{fig:bsf_scatter_error1} but for the noisier, validation set consisting of $71$ ZTF-like systems from \citet{Goldstein:2018bue}. Both  panels again show fits matching the true values. Uncertainties are 
somewhat larger compared to Fig.~\ref{fig:bsf_scatter_error1} due to a 
higher overall noise level in this set. 
} 
\label{fig:more_sys_zoom2}
\end{figure}

Figure~\ref{fig:more_sys_zoom2} shows the results for the time delays 
and magnifications. The method holds, though on this set it gives a 
larger fraction of systems not accepted as having confident 
detection of unresolved lensing ($\dtbf<10$ days).

\begin{figure}
\centering
\includegraphics[width=0.4835\textwidth]{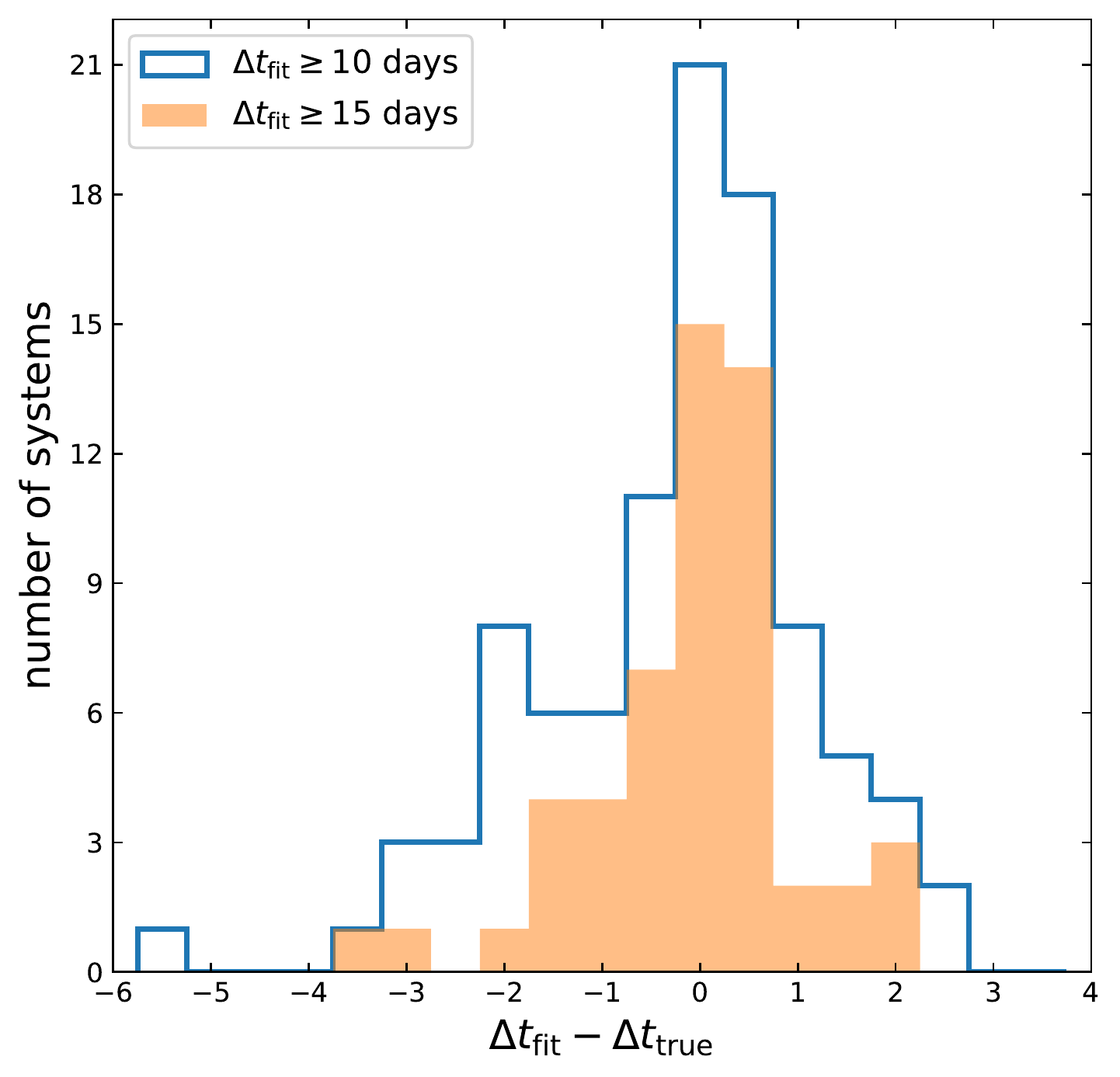}
\includegraphics[width=0.487\textwidth]{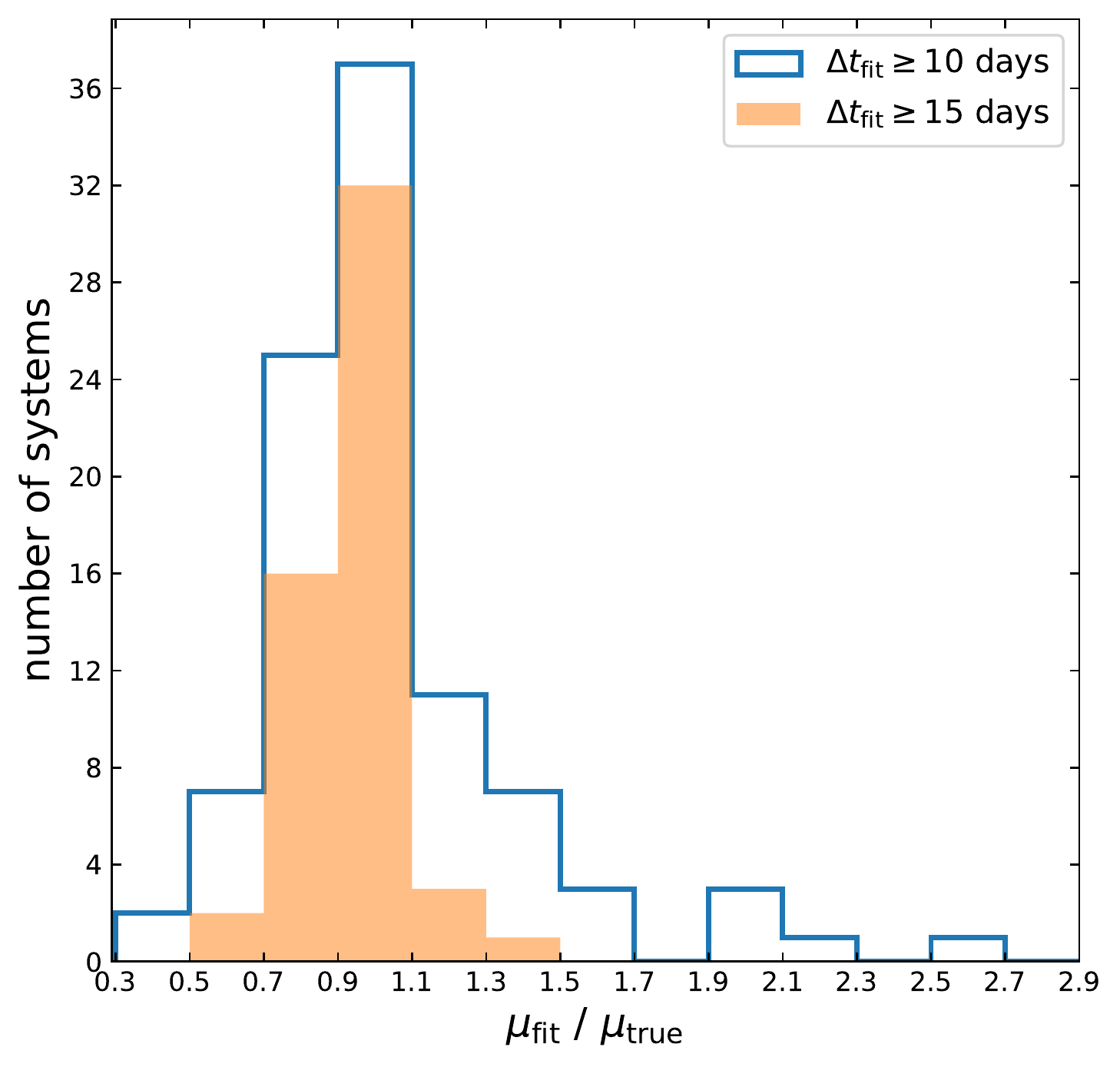}
\caption{As Fig.~\ref{fig:bsf} but for the consolidated set consisting of $33+71=104$ ZTF-like systems from \citet{Goldstein:2018bue}. Note that $97$ systems satisfy the condition $\dtbf \geq 10$ days that corresponds to the blue histogram in either of the panels whereas the orange histograms enclose $54$ systems. Again we find that for most of the systems our fit estimations are close to the respective true values. 
}
\label{fig:bsf_tot}
\end{figure}

Combining together the Goldstein sets of 33+71 lensed SN, 
Figure~\ref{fig:bsf_tot} presents histograms of scatter of the best fit 
from the true values, again for the outcomes where we accept the fit 
results -- either $\dtbf>10$ or 15 days. The histograms remain well 
peaked near the true values, though with increased scatter from the 
increased noise.

\subsection{Assessing Precision and Accuracy} \label{sec:accu}  

An important aspect of the use of lensed SN for cosmology 
is quantifying the fit precision and accuracy. We assess 
these below for the unresolved lensed SN we have studied. 

\begin{figure}
\centering 
\includegraphics[width=0.55\textwidth]{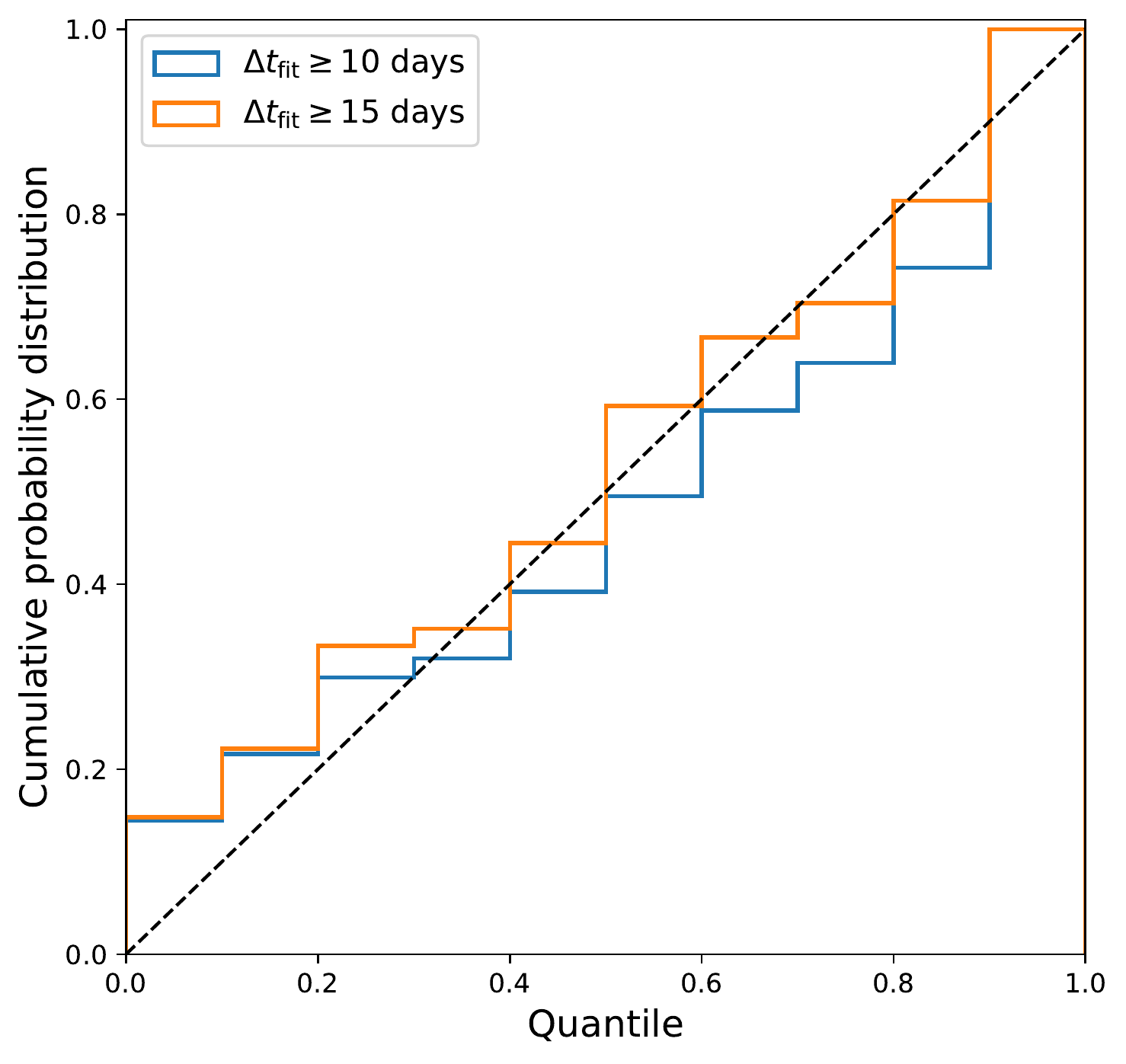}
\caption{Cumulative probability distribution from the consolidated set consisting of $33+71=104$ ZTF-like systems from \citet{Goldstein:2018bue}. The histograms basically show the fraction of the systems for which the true time delay falls within a given probability quantile of the fit distribution. Significant deviations from the diagonal dashed line 
indicate under- or over-estimation of the uncertainty. 
The amount of probability between the $16\%$--$84\%$ percentiles is found to be $48.5\%$ for the systems with $\dtbf \geq 10$ days (blue curve), and $59.3\%$ for the systems with $\dtbf \geq 15$ days (orange curve). Therefore, we find that the uncertainty in the estimated time delay, $\sigma(\dtbf)$, is underestimated by a factor of about $0.71$ for the blue curve and $0.87$ for the orange curve. 
This can be 
due in part to insufficient statistics, i.e.\ only 97 (54) systems in the blue (orange) histogram. 
}
\label{fig:cpd_tot}
\end{figure}

For  the fit precision, i.e.\ the uncertainty on the 
estimation, we investigate the cumulative pull 
distribution from the sampling chains, shown 
in Fig.~\ref{fig:cpd_tot}. By examining the amount of 
probability between the 16\%--84\% quantiles, 
we find that the chains are underestimating 
the $\dtbf$ uncertainties by approximately a factor of 
$0.71$ for the systems with $\dtbf\geq 10$ (and $0.87$ for $\dtbf \geq 15$ days). This is not uncommon for flexible model spaces 
where the full parameter space is not thoroughly 
sampled. 
Indeed, as mentioned in Sec.~\ref{sec:priorhyp} some of 
the priors on the non-physical hyper-parameters are 
informative, i.e.\ running into the boundary (and 
extending the boundary lowers the code efficiency and 
convergence). Thus the details of the derived probability 
distribution functions (PDFs), including potentially for the 
time delays and magnifications, can be affected. However, 
this does not seem to be the case for the best fit values, 
as seen in the accuracy tests below. 
Methods for improving uncertainty quantification 
are being developed in further work, but in this article 
we will not use the uncertainties quantitatively. 
We can still assess the accuracy of the method for best fits however.

For the fit accuracy, we want to ensure this in two respects. 
The first is a low fraction of false positives: 
cases that are fit as unresolved lensed SN even though in reality they  
are single unlensed images. The second is bias in the time delay 
measurement: are estimates statistically scattered about the truth or 
are they systematically offset low or high. 

The false positive rate was estimated by running 360 unlensed SNe in Sec.~\ref{sec:unlens} in addition 
to the 12 in Sec.~\ref{sec:3a}, as well as 
including unlensed SNe in the blind data 
set in  Sec.~\ref{sec:blind}, and considering what fraction of their fits have 
spurious time delay measurements inconsistent with zero delay. 
From the results in those subsections, the 
fraction of false positives can be estimated as $\lesssim5\%$ (19 out of 381), mostly due to the higher noise levels, if we use $\dtbf>10$ days, 
but 0 out of 381 for $\dtbf>12$ days. 

The second type of accuracy is bias. 
To assess bias, we compare the ensemble of fit time delays to the true 
inputs through \cite{Liao:2014cka,Hojjati:2014yta} 
\be 
A=\frac{1}{N_{\rm fit}} \sum \frac{\dt_{\rm fit}-\dt_{\rm true}}{\dt_{\rm true}}\,. \label{eq:A}
\ee  
Note that this uses the best fit and does not involve the fit 
uncertainty. 
Fits that scatter positive and negative about the truth 
for the time delay do not bias cosmology estimation; only a 
systematic offset relative to truth shifts the cosmology\footnote{ 
This holds for time delay distance; parameters related nonlinearly, 
such as the Hubble constant, can still be biased as high fluctuations 
do not perfectly balance low fluctuations if the scatter is  
not small, as for any probe.}. 

\begin{figure}
\centering
\includegraphics[width=0.55\textwidth]{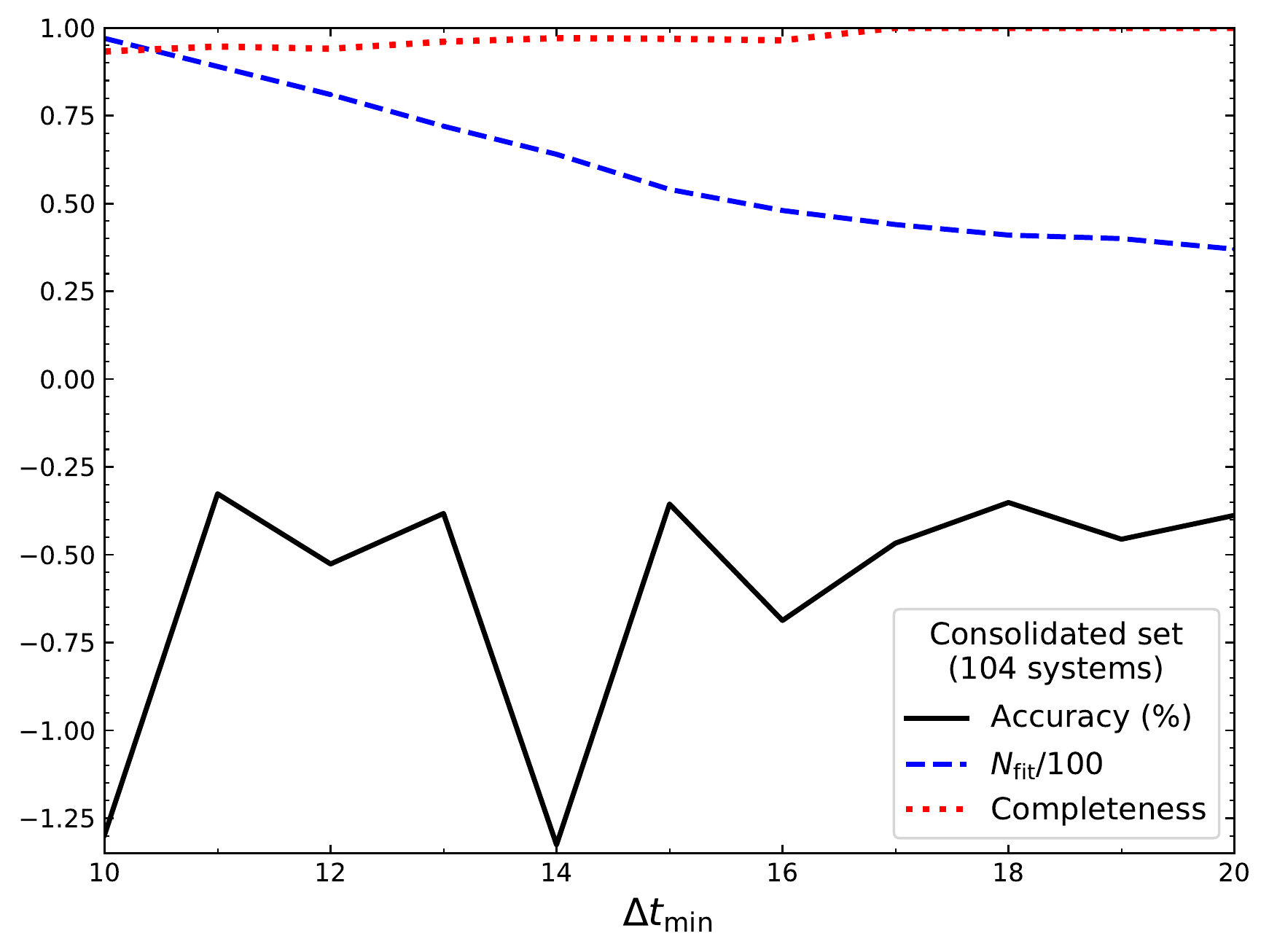}
\caption{
The accuracy $A$, Eq.~\eqref{eq:A}, important for using time delays as  a robust 
cosmological probe, is plotted vs ${\Delta t}_{\rm min}$, where 
only systems with $\dtbf \geq {\Delta t}_{\rm min}$ are accepted 
as unresolved lensed SN. The 
accuracy mostly lies around 
0.3\%--0.5\%. For limited numbers of systems in the sample, deviations 
in one or two systems can impact the accuracy; the number of accepted 
systems is shown by the dashed blue curve $N_{\rm fit}/100$. The 
completeness, i.e.\ the 
fraction of systems that should have been accepted given a perfect 
fit, and were accepted as robust unresolved lens SN, is shown by the dotted red curve. 
}
\label{fig:Aest}
\end{figure}

Figure~\ref{fig:Aest} shows the accuracy as a function of which 
fits we accept. Nominally, from previous sections, we have confidence  
in results with $\dtbf\ge10$ days, i.e.\ $\dt_{\rm fit,min}=10$ days. 
We study the results as a function of $\tmin$, finding that as soon 
as we move away from the boundary of 10 days, where slight variations 
in fit can cause the best fit to move above or below $\tmin$ despite 
much of the probability distribution remaining below, the accuracy 
ranges mostly over the 0.3\%--0.5\% level. For cosmology use with a 
rather advanced data set with small statistical uncertainty, 
\citet{Hojjati:2014yta} derived a  0.2\% accuracy requirement --  
that was for lens systems giving 1\% distance measurements in each of 
six lens redshift bins from $z_{\rm lens}=0.1 \text{--} 0.6$ (see also~\cite{Aghamousa:2014uya,Aghamousa:2016mvk}). Since that is 
rather ambitious for unresolved lensed SNe alone, we regard  $\sim0.5\%$ 
accuracy as quite good enough to start.

One might also consider false negatives, where the fit considers 
multiple images as one  unlensed source. We will mostly address this in future work 
where we fit for the number of images, e.g.\ does the algorithm 
accurately identify the number of components. Here we simply mention 
that for systems with true time delays more than ten days, there are 
zero cases where the time delay fit is consistent with zero delay, 
i.e.\ no lensing. We can also quantify the completeness, i.e.\ how 
many systems did we have sufficient confidence in the fit to include 
in the results. We quantify this by taking the the total number with  
$\dtt\ge\Delta t_{\rm min}$ and asking how many of these did we get any fit for 
(i.e.\ $\dtbf>10$ days). The completeness is shown as the dotted red 
curve in  Fig.~\ref{fig:Aest} and ranges from 93\% at $\Delta t_{\rm min}=10$ days 
to 97\% at $\Delta t_{\rm min}=14$ days and 100\% at $\Delta t_{\rm min}=17$ days.

\section{Conclusions} \label{sec:concl} 

Strong gravitational lensing allows us to see a source in different positions and at different times, and so can serve as an important cosmological probe. Lensed 
SN have advantages over lensed quasars in having a better understood 
intrinsic light curve with a characteristic time scale of order 
a month, allowing time delay estimation after months rather than 
many years of monitoring. Surveys underway, such as ZTF, and 
forthcoming, such as ZTF-2 and LSST, will greatly increase the number 
of lensed SN detected. However, if the time delay is of order or 
shorter than the characteristic SN timescale, i.e.\ about a month,  
then the lensed nature may not be obvious either from the observed 
lightcurve (a superposition of the image lightcurves) or spatially 
resolved on the sky. 

Here we investigate these unresolved lensed SN, and put forward a 
method for robustly extracting the time delays (and image 
magnifications). The time delays are directly related to the time 
delay distance of the supernova, adding a highly complementary  
cosmological probe to standard luminosity distance measurements. 
Knowledge of the magnifications  provides important constraints on 
the lens modeling, needed as well. Once a transient is identified 
as a lensed supernova, additional resources such as high resolution 
imaging from ground based adaptive optics systems or space based 
telescopes can target it to resolve images spatially. 

Our method, involving an orthogonal basis expansion around a 
trial lightcurve, sampled over many parameters for each band and 
image through Hamiltonian Monte Carlo, shows great promise. We 
recover the time delays (and magnifications) accurately down to 
$\dtbf\gtrsim10$ days for sampling cadence and noise levels 
representative of ZTF and LSST surveys. The false positive rate 
(interpreting an unlensed SN as a lensed one) is below 5\%, and 
the completeness (finding actually lensed SN as lensed) is 
93\% or better. Time delay measurement can accurately control bias 
to $\sim0.5\%$. 

We emphasize that the fundamentals of this technique are broadly 
applicable and do not rely fundamentally on SN properties. The method 
can be used for any transient lightcurve, with proper adjustment 
of hyperparameter priors. We have not yet tried this. Hamiltonian 
Monte Carlo provides a potentially robust sampling technique for 
the multidimensional parameter space necessary; our particular 
cases allowed 24 parameters but there is no innate limit. 

Significant further developments can be made. Our future work will 
include fitting for the number of images as part of the process 
(here we did only 1 vs 2, i.e.\ unlensed vs two lensed images), 
increasing flexibility in the template 
plus basis expansion approach, accounting for microlensing, and 
establishing cadence, noise, or other requirements to push the 
minimum time delay for which we have confidence in the results 
into the regime $\dtbf<10$ days. And of course we greatly look 
forward to applying the method to real data from surveys and 
discovering a significant sample of unlensed SN.

\acknowledgments 
SB thanks Ayan Mitra, Tousik Samui, and Shabbir Shaikh for their crucial helps at different stages of the project.
AK and EL are supported in part by the 
U.S.\ Department of Energy, Office of Science, Office of High Energy 
Physics, under contract no.~DE-AC02-05CH11231. EL is also supported under DOE Award DE-SC-0007867 and by 
the Energetic Cosmos Laboratory.  
AS would like to acknowledge the support of the Korea Institute for Advanced Study (KIAS) grant funded by the government of Korea.

\appendix  

\section{Hyperparameter and time delay priors} \label{sec:apxcode} 

The priors on the intrinsic lightcurve parameters, the hyperparameters, 
are presented in Table~\ref{tab:priors}. Here we give some more detail. 
We again caution that priors tend to involve the astrophysics of the 
source, and their specific values 
need to be adjusted depending on the type of transient. 
Here we focus on normal Type Ia supernovae. Apart from the specific 
priors, however, the method described in the main text should be 
broadly applicable to many unresolved lensed transients. 

The parameter $b_j$ (known as the scale parameter in lognormal 
distributions) is a logarithmic time variable that gives the median of the distribution. Since a SN falls more 
slowly than it rises, and it peaks about 20 days after explosion, 
the median is around 30 days. This translates to $b_j\approx 3.4$. 
Less than 20 days or longer than 60 days (e.g.\ for the longer wavelength 
bands) seem to be  unuseful parts of the parameter space, so we adopt 
$b_j\in[3.0,4.1]$. In the fits to the data, we find $b_j\approx 3.4$ 
and none of the fits approach closely the prior bounds. 

The parameter $\sigma_j$ (known as the shape parameter) relates to the (logarithmic) width of the distribution and interacts with $b_j$ to 
give skewness, shifting the mode (and mean) from the median. In 
particular the mode, i.e.\ maximum flux, occurs at $\exp(b_j-\sigma_j^2)$. 
If the median is 30 days and the mode is at 20 days after explosion, 
then $\sigma_j\approx0.64$. Extending the maximum all the way to 30 days 
after explosion would still give $\sigma_j<0.8$ for the largest $b_j$, 
i.e.\ the latest median. For the maximum no earlier than 17 days after 
explosion, one gets $\sigma_j>0.4$ for the smallest $b_j$. Thus the 
range adopted is $\sigma_j\in[0.4,0.8]$. 
In the fits to the data, we find $\mu\approx 0.64$ 
and none of the fits approach closely the prior bounds. 

Note these priors are specific to normal Type Ia supernovae we would 
use for cosmology. 
They are motivated by the lognormal template, which will  be further modified by the basis expansion terms. 
However, they work well in practice. 
For other supernovae, or other transients, the  
priors would need to be adjusted although the fitting method should 
still be substantially unchanged. 

The fit shift parameter $t_1$ comes from numerics using 
noisy data, not the astrophysics of supernovae. For low flux, photon 
noise and photometry scatter are especially influential, and the 
explosion time can be misestimated. Introducing $t_1$ helps mitigate 
this as the fit can be shifted to be optimized taking into account 
the full run of data. Numerical experimentation shows that 
$t_1\in[-5,5]$ days provides a reasonable estimation, though this parameter is less well constrained. 

For the normalization prior $N_j$, this is a convolution of the 
intrinsic lightcurve amplitude, the scaling adjustment by the 
basis function multiplication (essentially $C_{0,j}$), and the 
(not directly observed) magnification of the first image. Therefore 
it is a combination of supernova and lensing properties, and 
one should also keep in mind numerical efficiency. This can become complicated, 
but in brief we set priors on $N_j$ depending on the single/double 
peak nature of the observed lightcurve, taking into account the 
observed maximum flux, and the physically reasonable range of 
magnifications. 
See Appendix~\ref{sec:apxnorm} for further details. 

The time delay fit parameter itself has a prior as well, as we 
described in Sec.~\ref{sec:dtprior}. Here we go into additional  
detail. In the event of a single flux peak, if its width 
(full width at $80\%$ peak height, $w_{80}$) is more than 20 days 
we shift the prior on $\dt$. Taking $\delta t = (w_{80}-20)/1.8$, 
the prior on $\dt$ becomes $\delta t \leq \dt \leq 30+\delta t$. 
In addition, since the relative magnification cannot be very small 
in these cases, we impose a lower limit on $\mu$ leading to the prior 
$0.25 \leq \mu \leq 4$. In the two peak case there is no need for 
adjustment of the priors. Again, these priors are specific to normal 
Type Ia supernovae, and should be adjusted for other transients.

\section{Prior on the normalisation parameter} \label{sec:apxnorm} 

Separation of the basis expansion factor $\czj$ within 
the normalization factor $N_j$ in Eq.~\eqref{eq:template} 
can improve convergence in the sampling. 
We write 
\be
\label{eq:Nj}
N_j=C_{0,j} {\mathcal A}_j \, \exp(b_j-\sigma_j^2/2)\;. 
\ee 
The exponential term ensures that the flux function in 
Eq.~\eqref{eq:template} without shape modifications from the basis expansion terms, 
\bea
\label{eq:final} 
\F_j(t)&=&{\mathcal A}_j C_{0,j}\, \exp(b_j-\sigma_j^2/2) \left( \frac{1}{t} \right) \exp\left[-\frac{(\ln t - b_j)^2}{2 \sigma_j^2}  \right]\;, 
\eea
has a maximum height ${\mathcal A}_j \czj$, independent of $b_j, \sigma_j$. 
Now 
$C_{0,j}$ is used as the fit parameter. The amplitude  ${\mathcal A}_j$ is determined from the observed data as a fraction of the maxima of the observed summed light curve,
\begin{equation}
{\mathcal A}_j=\frac{1}{k} \times {F_j}^{\rm max}\;,
\end{equation}
where the constant $k$ can be any number greater than unity. 

Now suppose the true (unobserved) maxima of the light curves of the first and the second image are $B_{j1}$ and $B_{j2}$ respectively in the $j$th filter. 
If we set $\czj$ such that ${\mathcal A}_j \czj \approx B_{j1}$ then  
the desired value of $\czj$ would be
\be
\czj=\frac{B_{j1}}{{\mathcal A}_j}=\frac{k B_{j1}}{{F_j}^{\rm max}}\;.
\ee
We have two inequalities between $B_{j1}, B_{j2}$, and  $\Fm$:
\begin{itemize}
    \item $B_{j1} \leq \Fm$ which yields $\czj \leq k$.
    \item $B_{j1} + B_{j2} \geq \Fm$ which yields $\czj \geq k/(1+\mu)$ given that the true magnification is $\mu=B_{j2}/B_{j1}$.
\end{itemize}
Therefore, 
a useful prior on $\czj$ is 
\be
\frac{k}{1+\mu} \leq \czj \leq k\;.
\ee 
Taking a wide prior on $\mu$ as $0 \leq \mu \leq 4$, 
and setting $k=2$ (as if the two images contribute equally, i.e.\ not favoring either), the prior on $\czj$ becomes
\be
0.4\leq \czj \leq 2\;. 
\ee 
This is what we list in 
Table~\ref{tab:priors} and use in the text.

\bibliography{unressn}

\end{document}